\documentclass[%
 reprint, prd, 
superscriptaddress,
 amsmath,amssymb,
 aps,
]{revtex4-2}

\usepackage{graphicx}
\usepackage{dcolumn}
\usepackage{bm}
\usepackage{hyperref}

\begin{document}

\preprint{APS/123-QED}

\title{Effective Quantum Gravitational Collapse in a Polymer Framework}

\author{Lorenzo Boldorini}%
 \email{boldorini.1843532@studenti.uniroma1.it}
\affiliation{Department of Physics, “La Sapienza” University of Rome, P.le Aldo Moro 5, 00185 Rome, Italy}
\author{Giovanni Montani}
\email{giovanni.montani@enea.it}
\affiliation{ENEA, Fusion and Nuclear Safety Department, C. R. Frascati,\\Via E. Fermi 45, 00044 Frascati (RM), Italy}
\affiliation{Department of Physics, “La Sapienza” University of Rome, P.le Aldo Moro 5, 00185 Rome, Italy}

\date{\today}

\begin{abstract}
 We study how the presence of an area gap, different than zero, affects the gravitational collapse of a dust ball. The implementation of such discreteness is achieved through the framework of polymer quantization, a scheme inspired by loop quantum gravity (LQG). We study the collapse using variables which represent the area, in order to impose the non-zero area gap condition. The collapse is analyzed for both the flat and spherical Oppenheimer-Snyder models. In both scenarios the formation of the singularity is avoided, due to the inversion of the velocity at finite values of the sphere surface. This happens due to the presence of a negative pressure, with origins at a quantum level. When the inversion happens inside the black hole event horizon, we achieve a geometry transition to a white hole. When the inversion happens outside the event horizon, we find a new possible astrophysical object. A characterization of such hypothetical object is done. Some constraints on the value for the area gap are also imposed in order to maintain the link with our already established physical theories.
\end{abstract}

\maketitle


\section{\label{sec:introduction}INTRODUCTION}
One of the most notorious features of the theory of General Relativity (GR) is the fact that it predicts the formation of black holes during gravitational collapse of stars\cite{Misner:1973prb}\cite{Weinberg2008}. The theory predicts that when the internal pressure of a star is not strong enough to counterbalance the gravitational pull of its mass, the star collapses to become a black hole. Sometimes this collapse could be halted by an internal pressure with quantum origins: electron degeneracy pressure is responsible for the formation of white dwarfs, while neutron degeneracy pressure, with the aid of the strong force, is responsible for the formation of a neutron star. More details can be found in \cite{Shapiro1983-fp}.\\
This is a remarkable result to keep in mind for the scope of this paper: a quantum effect induces a pressure which has macroscopic consequences.\\
When the mass is too high the internal quantum pressure is not enough to prevent the formation of a black hole. The mass range after which collapse is inevitable is called Tolman-Oppenheimer-Volkoff Limit, the most recent theoretical esteem being $2.2-2.9 M_\odot$\cite{Kalogera_1996}, while constraints from gravitational wave signals put the limit at $2.01-2.16 M_\odot$\cite{Rezzolla_2018}.\\
When such limits are passed, the collapsing solution present also one of the most critically discussed points of the theory: the singularity.\\
Singularities are mostly though to be a sign of incompleteness for the theory of GR, invoking for the search of a more complete theory which encompass quantum mechanics. The first approach to achieve the quantization of gravity was the Wheeler-DeWitt formulation \cite{WdW67}. This approach was rather conservative, since it was based off canonical quantization of a constrained system: gravity.\\
The evolution of this canonical quantization procedure is today called Loop Quantum Gravity (LQG), a canonical quantization scheme based on the Ashtekar-Barbero-Immirzi's connection variables. This approach more closely resembles the usual gauge theories encountered in theoretical physics, but possessess also very peculiar features, such as a discrete spectrum for the operator which represents the area, or like a diffeomorphisms invariant Hilbert space. A very detailed account for such a quantization procedure for gravity is given in \cite{Thiemann_2007}.\\
Our analysis in this work is based on such a canonical approach. We will adopt the Oppenheimer-Snyder\cite{Oppenheimer_Snyder_1939} model (OS) for gravitational collapse, which is the first model developed to describe the gravitational collapse of a sphere without internal pressure.\\
This model is described internally by an evolving FLRW metric, while outside the geometry is fixed to be the Schwarzschild spacetime. Our quantization procedure is applied then to the internal dynamical FLRW metric.\\
We make use of the so called polymer quantization\cite{Corichi_2007}, a canonical quantization scheme which implements the non-vanishing area gap and the diffeomorphisms invariance of the Hilbert space, like in the LQG framework.\\
The main difference is that the scale factor $a(t)$ of the FLRW solution does not act as a field but rather as a single degree of freedom, so the polymer technique is to be understood as a LQG-like quantization for a single degree of freedom.\\
We will build an effective Hamiltonian for the OS model where it gets corrections from the polymer quantization procedure. The meaning of such a method of analysis is that we consider the classical trajectories to be the trajectories for the expectation values of the quantum operators, which we assume to be peaked around the classical trajectory.\\
Our aim in this paper is to understand whether or not the implementation of the quantum nature of gravity could prevent the formation of the above mentioned singularity. This phenomenon is already known in the canonical quantization approach for the collapse of shells \cite{sing_kief} and for the Lemaitre-Tolman-Bondi (LTB) model \cite{LTB_Kief}\cite{Malafrina}.\\
As main issue of our study we demonstrate that the collapse singularity is removed in favor of a possible guess on the existence of a quantum black-to-white hole transition.\\
Furthermore, as most interesting phenomenological outcome, we see that under suitable conditions, but independent of the object mass, the collapse is unable to cross the horizon. As a result, we deal with a new hypothetical astrophysical object, which could also be over-critic but radially oscillating between two super-horizon configurations.\\
Finally we show how the emergence of the polymer phenomenology can be interpreted as a negative quantum pressure, which form is explicitly determined. \\
The paper is organized as follows. In Section \ref{sec:ham_OS} we first give a review of the Hamiltonian formulation of the Oppenheimer-Snyder model, then we introduce a new pair of canonical variables and we use them to study the classical model. In Section \ref{sec:eff_OS} we introduce the effective model, built using the polymer corrections, and solve the equations of motion for both the area and momentum. Some relevant aspects are discussed after presenting each solution. In Section \ref{sec:phen} the phenomenology of the effective model is discussed: we will characterize a kind of quantum gravitational pressure generated by polymer quantization, and macroscopic and quantum effects will be analyzed.\\
Section \ref{sec:conc} closes the paper with a summary, some remarks and future perspectives for this line of research.\\
Throughout all this paper we will use geometrized $(G=c=1)$ units.
\section{\label{sec:ham_OS}HAMILTONIAN OPPENHEIMER-SNYDER MODEL}
In this section we are going to analyze the geometry behind the gravitational collapse model.\\
First we will introduce the solution found by Oppenheimer and Snyder\cite{Oppenheimer_Snyder_1939}, then we will see how an Hamiltonian description of the very same model can be formulated \cite{Gourgoulhon:2007ue} \cite{ADM_1959} \cite{Poisson_2004}, giving rise to a constrained model. \\
We will later introduce a new set of canonical variables, well suited for the effective analysis to be made in the next section. This new variables will then be used to solve the classical model, in order to compare it to effective one developed in section \ref{sec:eff_OS}.
\subsection{\label{subsec:Opp-Sny}The Oppenheimer-Snyder Geometry}
Let's start our analysis with the description of the Oppenheimer-Snyder model geometry.\\
The spacetime manifold is composed of mainly three pieces\cite{Hájíček_Kijowski_1998}: an interior region, where a dust is present, an exterior vacuum region and at last a boundary, which will be set to infinity.\\
The manifold, denoted by $\Omega$, is then split into two portions: the matter region $\Omega_-$ and the vacuum region $\Omega_+ = \Omega\;\backslash\;\Omega_-$. The timelike boundary between the two portions of the manifold is denoted by $\Gamma$ while the boundary of the manifold is denoted by $\Gamma_+$ .\\
We are willing to give an Hamiltonian description of the model, so we need to apply the same subdivision also to the spacelike hypersurfaces.\\
Denoted the spacelike hypersurfaces by $\Sigma_t$, the spacelike interior matter region is given by $S_- = \Omega_- \cap \Sigma_t$ while the exterior vacuum is given by $S_+ = \Omega_+ \cap \Sigma_t$. The boundary between the two regions is where the spatial surface of the dust sphere lies, given by $\Sigma_S = \Gamma \cap \Sigma_t$.\\
The boundary of the manifold is given by $\Sigma_+ = \Gamma_+ \cap \Sigma_t$.\\
Since we are dealing with a spherical symmetric distribution of matter, the geometry of the spacelike hypersurface is well described by the set of spherical coordinates $x^i = (r,\theta, \phi)$ where $r\geq 0$ has the meaning of a radial coordinate, with $r=0$ in the center of the dust ball. The boundary surface $\Sigma_S$ is held fixed at a finite constant coordinate radius $r=r_S$ while the boundary of the manifold $\Sigma_+$ is placed at a coordinate radius of $r=r_+$.\\
It is very important to note that the coordinates are not observables here, since observables are given by functions on the phase space. The coordinates thus are just a parametrization of the manifold, useful to relate canonical variables.\\
Let's see now how the geometry in each manifold portion is described.\\
The geometry inside the region $\Omega_-$ is determined by the Friedmann-Lemaître-Robertson-Walker (FLRW) metric\cite{Misner:1973prb}:
\begin{equation}
\label{eqn:flrw_met}
ds^2_{(-)} = -d\tau^2 +a^2(\tau)\left(\frac{dr^2}{1-\epsilon r^2} +r^2d\Omega^2\right)
\end{equation}
Where $\epsilon \in \{ 0, +1,-1\}$ respectively for flat, spherical and hyperbolic $\Omega_-$. The dimensions are such that $\big[\epsilon\big] = L^{-2}$ and $a(\tau)$ is dimensionless.\\
$a(\tau)$ is the scale factor, while the coordinate $\tau$ is the proper time measured by an observer comoving with the dust in $\Omega_-$.\\
The flat geometry describes collapse starting from infinity with zero velocity, the hyperbolic geometry describes collapse starting from infinity with a given initial velocity and, at last, the spherical geometry describes collapse starting from an initial finite radius with zero initial velocity\cite{PG_coord_Kanai}\cite{Casadio_1998}.\\
The geometry inside the region $\Omega_+$ is determined by the Birkhoff theorem and is the Schwarzschild spacetime geometry\cite{Misner:1973prb} given by:
\begin{equation}
\label{eqn:sch_met}
ds^2_{(+)} = -F(R)dT^2 + F^{-1}(R)dR^2 +R^2 d\Omega^2
\end{equation}
The function $F(R)$ is given by the usual:
\begin{equation}
\label{eqn:sch_F}
F(R) = 1- \frac{2M_{S}}{R}
\end{equation}
Where $M_{S}$ is the dust sphere mass, namely the ADM mass of the system.\\
The vector $\partial / \partial T$ is a Killing vector of the line element (\ref{eqn:sch_met}) and it's orthogonal to the constant $T$ hypersurfaces, while $R$ is the radial Schwarzschild coordinate.\\
From now on will only cover the flat and spherical cases, since they appear to be the most interesting from an astrophysical point of view.\\
\subsection{\label{subsec:Hami_mod}Hamiltonian Model}
Let's turn our attention now to the Hamiltonian description of said model, which is portrayed in \cite{Kiefer_Mohaddes_2023}\cite{Schmitz_2020}.\\ 
The dust model adopted is the one provided by  Brown and Kuchař\cite{Brown_Kuchař_1995} of an isotropic, homogeneous pressureless dust. The canonical variable $\tau$ describes the dust proper time, while its conjugate momentum $P_\tau$ is the energy inside the dust sphere.\\
Before the derivation of the other canonical variables we define the useful quantity:
\begin{equation}
\begin{split}
      V_S &=  \int_0^{r_S} dr \frac{r^2}{\sqrt{1-\epsilon r^2}} =\\
      &= \begin{cases}
r_S^3 / 3&\epsilon = 0\\
\frac{-\sqrt{\varepsilon r_S^2 - r_S^4 \varepsilon ^2}+ \text{arctan}\left(\frac{\sqrt{\varepsilon r_S^2}}{\sqrt{1-\varepsilon r_S^2}-1}\right)+\frac{\pi}{2}}{\varepsilon^{3/2}}  &\epsilon = +1\varepsilon 
    \end{cases}
\end{split}
\end{equation}
Where $\varepsilon \equiv 1[L]^{-2}$ for dimensional reasons. This quantity is such that $4\pi V_S$ is the parametric volume of a ball of coordinate radius $r=r_S$ in the two geometries.\\
We also identify the lapse function to be $N_\tau^2(t) = \left(\frac{d\tau}{dt}\right)^2$.\\
Defining $\rho$ as the dust energy density, the momentum $P_\tau$ is given by:
\begin{equation}
    P_\tau = 4\pi V_S a^3(t)\rho
\end{equation}
The canonical variables which describe the interior geometry are the scale factor $a$ and its conjugate momentum $P_a$, while the exterior geometry is described by the canonical variables $(R,T)$, the Schwarzschild coordinates, and their conjugated momenta $(\tilde{P}_R, P_T)$.\\
To obtain the Hamiltonian description of the model the following canonical transformation is made:
\begin{equation}
    R_s=ar_S \qquad P_s = P_a/r_S
\end{equation}
Those are the proper radius and momentum of the surface of the dust cloud.
Having also defined:
\begin{equation}
\Xi = \frac{r_S^3}{3V_S}
\end{equation}
Which is the ratio between the volume of a flat sphere and the proper volume of a sphere of equal radius, we see that the relation between the sphere mass and the dust energy is given by:
\begin{equation}
    M_S = \Xi P_\tau
\end{equation}
The last ingredient needed to obtain an Hamiltonian formulation of the theory is a parametrization of the Schwarzschild time on the dust surface, labeled $T_S(t)$, which is required to make the action canonical. \\
Following \cite{Kiefer_Mohaddes_2023} we see that we have two choices.\\
For the flat case, i.e. $\epsilon = 0$, the function is given by:
\begin{equation}
\label{eqn:PG_time}
T_S^{PG}= t +\int dr \frac{\sqrt{2M_S/R}}{1-\frac{2M_S}{R}}
\end{equation}
Those are called Painlevé-Gullstrand\cite{Painleve_1921}\cite{Gullstrand:1922tfa} coordinates.
For the spherical case, i.e. $\epsilon = +1$, the function is given by:
\begin{equation}
\label{eqn:GH_time}
T_S^{GH} = \frac{1}{1-\frac{2M_S}{R}}\left(t +\int_{R_i}^R dy \frac{\sqrt{\frac{2M_S}{y} - \frac{2M_S}{R_i}}}{1-\frac{2M_S}{y}}\right)
\end{equation}
Where:
\begin{equation}
R_i = \frac{2M_S}{R^\phi_{\theta\phi\theta}}
\end{equation}
Those are called Gautreau-Hoffman\cite{Gautreau_Hoffmann_1978} coordinates. \\
For increased readability we define:
\begin{equation}
\kappa_S = \epsilon r_S^2
\end{equation}
Which is a dimensionless parameter.\\
We are now ready to write down the action for the Oppenheimer-Snyder model:
\begin{equation}
\begin{split}
\label{eqn:CANON_ACT}
&S = \int dt \left\lbrace\left( P_\tau \dot{\tau}  + P_s\dot{R}_s - H_-\right) \right.\\
&\left.+ \int_{r_S}^{\infty}dr\left[\left(\tilde{P}_R\dot{R} + P_T\dot{T}\right) - \left(\beta^T P_T +\beta^R  \tilde{P}_R\right)\right] \right\rbrace
\end{split}
\end{equation}
With the Hamiltonian:
\begin{equation}
\label{eqn:CANON_HAM_CLAS}
\begin{split}
&H_- =  N_\tau \mathcal{H}_-\\
&\mathcal{H}_- = -\frac{\Xi }{2}\left(\frac{P_s^2}{R_s} + \frac{\kappa_S }{\Xi ^2}R_s  -2 \frac{P_\tau }{\Xi }\right)  
\end{split}
\end{equation}
The constraints are:
\begin{equation}
\label{eqn:consfin_old}
\mathcal{H}_- \approx 0 \qquad P_T \approx 0 \qquad \tilde{P}_R \approx 0
\end{equation}
This means that the only dynamical degrees of freedom are $\tau$ and $R_s$, or more properly $a$ since $R_s$ is just one of the possible rescaled versions. The variables $(R,T)$ are then fully constrained, with no dynamics: the external geometry really is fixed to be Schwarzschild up to the diffeomorphisms generetad by the last two constraints in (\ref{eqn:consfin_old}).\\
\subsection{\label{subsec:area_varia}Area Variables}
We now aim to develop a new formulation in order to solve both the classical and the effective dynamics. We will build a new set of variables and use them to recast the Hamiltonian OS model from a different point of view.\\
We start by making a canonical transformation, adopting the new set of variables more suited for the effective model.\\
This new change of variables is made through the following canonical transformation and its inverse:
\begin{align}
&\begin{cases}
\mathcal{A}(R_s, P_s) = 4\pi R_s^2\\
P_{\mathcal{A}}(R_s, P_s) = \frac{P_s}{8\pi R_s}
\end{cases}\\
&\begin{cases}
R_s(\mathcal{A}, P_{\mathcal{A}})=\sqrt{\frac{\mathcal{A}}{4\pi}}\\
P_s(\mathcal{A}, P_{\mathcal{A}})=P_{\mathcal{A}}\sqrt{16\pi \mathcal{A}}
\end{cases}
\end{align}
Where $\mathcal{A}$ is the proper surface area of a sphere of proper radius $R_s$ and $P_{\mathcal{A}}$ is its conjugate momentum.\\
It is clear that $P_s \dot{R}_s = P_{\mathcal{A}} \dot{\mathcal{A}}$ so that the action in the new variables reads:
\begin{equation}
        \label{eqn:act_SQ}
             \begin{split}
&S =\int dt \left\lbrace\left(P_\tau \dot{\tau}  + P_{\mathcal{A}} \dot{\mathcal{A}} -N_\tau \tilde{\mathcal{H}}\right) \right. \\ 
&\left. +  \int_{r_S}^{\infty}dr\left[\left(\tilde{P}_R\dot{R} + P_T\dot{T}\right) - \left(\beta^T P_T +\beta^R  \tilde{P}_R\right)\right] \right\rbrace
    \end{split}   
\end{equation}
\\
With the new Hamiltonian given by:
\begin{equation}
\label{eqn:ham_SQ}
 \begin{split}
&\tilde{H}= N_\tau \tilde{\mathcal{H}} \\ 
&\tilde{\mathcal{H}} = -\frac{\Xi }{2}\left(32\pi^{\frac{3}{2}}P_{\mathcal{A}} ^2\sqrt{\mathcal{A}} + \frac{\kappa_S }{\Xi ^2} \frac{\sqrt{\mathcal{A}}}{\sqrt{4\pi}}  -2 \frac{P_\tau }{\Xi }\right) 
  \end{split}  
\end{equation}
\\
The constraints become:
\begin{equation}
\label{eqn:cosfin}
\tilde{\mathcal{H}} \approx 0 \qquad P_T \approx 0 \qquad \tilde{P}_R \approx 0
\end{equation}
\\
We will choose now the comoving gauge, which corresponds to the choice of $N_\tau = 1$, from which the following equations of motion are deduced:
\begin{align}
			&\dot{\mathcal{A}} = \frac{\delta \tilde{H}}{\delta P_{\mathcal{A}}} = -32 \pi ^{3/2} \Xi  P_{\mathcal{A}} \sqrt{\mathcal{A}}\label{eqn:A_clas_eom}\\
			&\dot{P}_{\mathcal{A}} = - \frac{\delta  \tilde{H}}{\delta \mathcal{A}} = \frac{\kappa_S }{8\Xi    \sqrt{\pi\mathcal{A}}} +\frac{8\Xi \pi ^{3/2} P_{\mathcal{A}}^2}{\sqrt{\mathcal{A}}}\label{eqn:PA_clas_eom}\\
			&\dot{\tau} = \frac{\delta  \tilde{H}}{\delta P_\tau} = 1\label{eqn:t_clas_eom}\\
			&\dot{P}_\tau  = -\frac{\delta  \tilde{H}}{\delta \tau} = 0\label{eqn:Pt_clas_eom}
\end{align}
We now start to develop the classical dynamics.
\subsection{\label{subsec:class_eqns}Classical Dynamics}
We now aim to solve the classical equations of motion in this new variables.\\
Equations (\ref{eqn:t_clas_eom}) and (\ref{eqn:Pt_clas_eom}) are easily solved, yielding:
\begin{align}
\tau(t) &= t \label{eqn:tau_tra}\\
P_\tau (t) &= const = \frac{M_S}{\Xi}\label{eqn:Ptau_tra}
\end{align}
Now we get onto the equations of motion for the area and its conjugate momentum.
\subsubsection{\label{subsubsec:clas_surf}Classical Surface Equation}
\begin{figure}[ht]
\centering
  \includegraphics[scale=0.6]{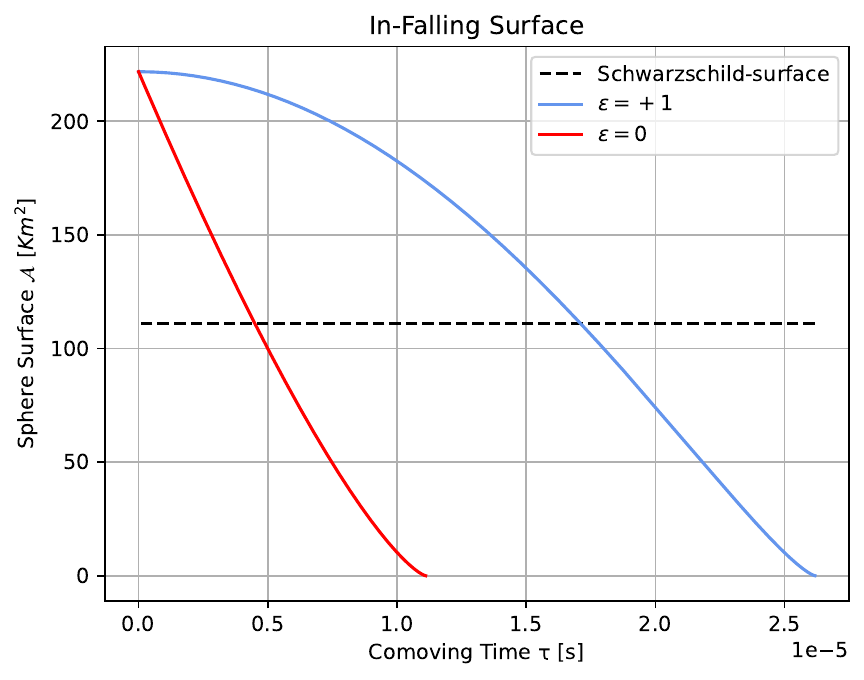}
  \caption{In-falling trajectories against comoving time $\tau$ for the two scenarios. The parameters chosen are $\alpha = 2$ and $M_S = M_\odot$.}
  \label{fig:infalling_classical_s}
\end{figure}

Starting from the one relative to the area, we first try to express the equation for the dust surface as only a function of its area. In order to do so we first evaluate $ P_{\mathcal{A}}^2(\mathcal{A})$ solving the constraint $\tilde{\mathcal{H}} \approx 0$, which gives:
\begin{equation}
\label{eqn:ham_con_cla_q}
P_{\mathcal{A}}^2 = \frac{4M_S\sqrt{\pi\mathcal{A}}  - \kappa_S\mathcal{A}}{64 \pi ^2 \Xi ^2\mathcal{A}}
\end{equation}
Extrapolating $P_{\mathcal{A}}$ from this and substituting it into $\dot{\mathcal{A}}$ we obtain the equations of motion:
\begin{equation}
\label{eqn:sdot_cla_pm2}
\dot{\mathcal{A}}^\pm = \pm 4 \sqrt{\pi }  \sqrt{4M_S  \sqrt{\pi\mathcal{A}}-\kappa_S\mathcal{A}}
\end{equation}
Here $\dot{\mathcal{A}}^-$ is the equation for the in-falling dust ball while $\dot{\mathcal{A}}^+$ is the one for the out-rising dust sphere.\\
Before solving this we seek for the maxima and minima of the canonical variable.\\
In order to achieve this, we impose the condition $\dot{\mathcal{A}}^2 = 0$, which we can solve easily for $\mathcal{A}$ after the substitution of $P_{\mathcal{A}}^2$ into $\dot{\mathcal{A}}^2$.\\
The solutions of the equation are:
\begin{equation}
\label{eqn:maxmin_cla}
\mathcal{A}_{max} = \frac{16\pi M_S^2}{\kappa_S^2} \qquad
\mathcal{A}_{min} = 0
\end{equation}
Here $\mathcal{A}_{max}$ is a maximum while $\mathcal{A}_{min}$ is a minimum, as shown by the domain restrictions on the equations of motion.\\
Analyzing the maximum will shed some light on the parameter $\kappa_S$: if we identify the maximum with the initial surface of the sphere, namely $\mathcal{A}_{max} = \mathcal{A}_0$, we see that the $\kappa_S = 0$ case is equivalent to saying that the maximum for the flat case is at $\mathcal{A}_{max} = +\infty$: this is consistent with the previous statement that $\epsilon=0$ is associated with the collapse from infinity, so the surface has no maximum value.\\
To study the spherical case we start by defining the Schwarzschild-surface as the surface of the event horizon:
\begin{equation}
\label{eqn:sch_surf}
\mathcal{A}_{SC} = 4\pi\left(2M_S\right)^2 = 16\pi M_S^2
\end{equation}
Defining also $\alpha$ as the ratio between the initial surface and the Schwarzschild-surface of the dust ball:
\begin{equation}
\alpha = \frac{\mathcal{A}_0}{\mathcal{A}_{SC}}
\end{equation}
We immediately see that the parameter $\kappa_S$ is related to the initial value of the surface via:
\begin{equation}
\label{eqn:epsparam}
\kappa_S =  \sqrt{\frac{1}{\alpha}} = \frac{2M_S}{R_{max}}
\end{equation}
Since for any physical situation $\alpha >1$, it means that $\kappa_S <1$ always.\\
The $\kappa_S = 0$ case then corresponds to the choice of $\alpha = \infty$, so the collapse starts at an initial radius infinitely far from the event horizon.\\
We now turn back to solve our equations of motion (\ref{eqn:sdot_cla_pm2}).\\
The equations are separable, so that they become:
\begin{equation}
 \frac{d\mathcal{A}^\pm}{ \sqrt{4 M_S  \sqrt{\pi\mathcal{A}}-\kappa_S\mathcal{A}}}=\pm 4 \sqrt{\pi } dt
\end{equation}
Those are two different sets of equations, two for the flat case and two for the spherical case. With the help of the integrals defined in Appendix \ref{app:integr}, and remembering that $t=\tau$, the solutions read:
\begin{align}
&\mathcal{A}^-:-4\sqrt{\pi} \;\tau = C^- + \mathcal{I}_\epsilon^{CA}(\mathcal{A}) \label{eqn:inf-cla}\\
&\mathcal{A}^+:4\sqrt{\pi} \;\tau = C^+ + \mathcal{I}_\epsilon^{CA}(\mathcal{A}) \label{eqn:ouris-cla}
\end{align}
We now need to obtain the constants of integration: for the in-falling branch we want that, at $\tau=0$, the surface has value $\mathcal{A}=\mathcal{A}_0$. This is also valid for the collapses coming from infinity: since they need to start at $\tau = -\infty$ we can choose a finite value for $\mathcal{A}$ when the comoving time is zero.\\
This means that:
\begin{equation}
C^- = -\mathcal{I}_\epsilon^{CA}(\mathcal{A}_0)
\end{equation}
For the out-rising solution, in order to have a continuous behaviour, we would like to join the two branches at the minimum $\mathcal{A}_{min}=0$. This is impossible because both branches are singular at $\mathcal{A}_{min}=0$.\\
If instead we take  $C^+ = \mathcal{I}_\epsilon^{CA}(\mathcal{A}=0^+)$ we obtain the time reversal of the collapse.\\
The solution then reads:
\begin{align}
&\mathcal{A}^-:-4\sqrt{\pi} \;\tau = -\mathcal{I}_\epsilon^{CA}(\mathcal{A}_0) + \mathcal{I}_\epsilon^{CA}(\mathcal{A}) \label{eqn:inf-cla_FIN}\\
&\mathcal{A}^+:4\sqrt{\pi} \;\tau =\mathcal{I}_\epsilon^{CA}(\mathcal{A}=0^+) + \mathcal{I}_\epsilon^{CA}(\mathcal{A}) \label{eqn:ouris-cla_FIN}
\end{align}
Clearly the time reversal is not a physical solution due to the presence of the singularity in the conjunction point.\\
We show in Fig. \ref{fig:infalling_classical_s} a plot of the classical in-falling trajectory in this new variables.
\\
Now its the time to solve the momentum equations.
\subsubsection{\label{subsubsec:clas_mom}Classical Momentum Equation}
\begin{figure}[ht]
	\centering
  \includegraphics[scale=0.6]{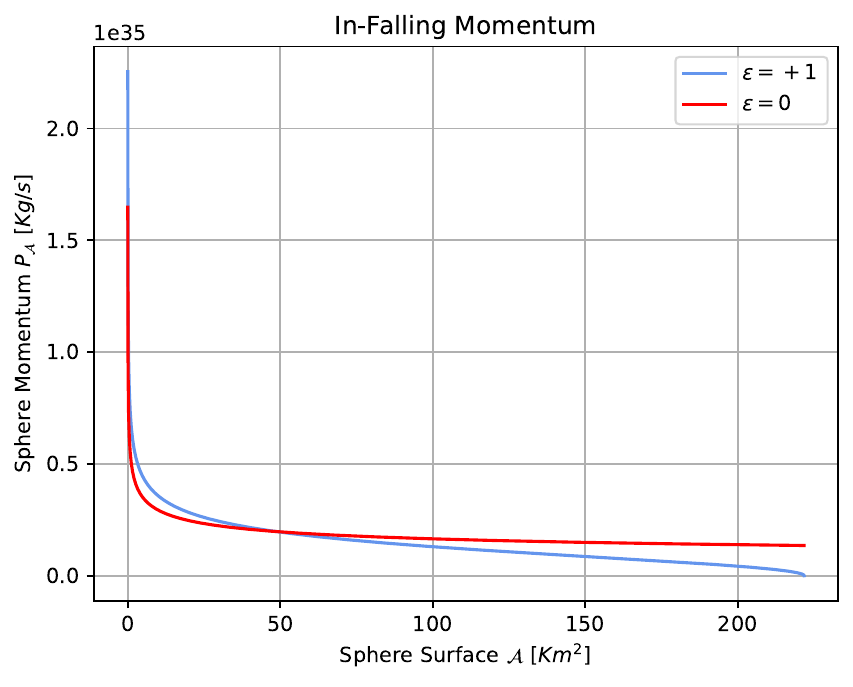}
  \caption{In-falling momentum trajectory for both scenarios. The parameters chosen are $\alpha = 2$ and $M_S = M_\odot$.}
  \label{fig:infalling_classical_q}
\end{figure}

We now solve the equation for the dust sphere momentum. \\
We will not solve the equation in function of the comoving time $\tau$ but solve for $P_{\mathcal{A}}$ as a function of $\mathcal{A}$. The reason why this appears more useful is because it gives the momentum when the sphere has surface $\mathcal{A}$, and since we have $\tau(\mathcal{A})$ we can evaluate $P_{\mathcal{A}}(\tau)$ from such relation. We have that:
\begin{equation}
P_{\mathcal{A}}':= \frac{dP_{\mathcal{A}}}{d\mathcal{A}}=\frac{\dot{P}_{\mathcal{A}}}{\dot{\mathcal{A}}} = -\frac{\frac{\kappa_S}{\pi ^2 \Xi ^2}+64 P_{\mathcal{A}}^2}{256 P_{\mathcal{A}}\mathcal{A}}
\end{equation}
Which yields the separable equation:
\begin{equation}
\label{eqn:eomQoS}
\frac{256 P_{\mathcal{A}} d P_{\mathcal{A}}}{\frac{\kappa_S}{\pi ^2 \Xi ^2}+64  P_{\mathcal{A}}^2} = -\frac{d\mathcal{A}}{\mathcal{A}}
\end{equation}
This is solved exactly with the aid of the integrals in Appendix \ref{app:integr}:
\begin{align}
&\epsilon = 0 \; : \qquad  P_{\mathcal{A}}( \mathcal{A}) =\frac{e^{-\frac{C_0}{4}}}{\sqrt[4]{\mathcal{A}}}\\
&\epsilon = +1 \; : \qquad  P_{\mathcal{A}}(\mathcal{A}) = \pm\sqrt{\frac{\frac{e^{-\frac{C_1} {2}}}{\sqrt{\mathcal{A}}}-\kappa_S}{64 \pi^2  \Xi^2 }}
\end{align}
The constants of integration are obtained by setting the momentum to be zero at the initial surface for the spherical case, while for the flat case, according to equations (\ref{eqn:A_clas_eom}) and (\ref{eqn:sdot_cla_pm2}), is recovered by setting at some given surface $\mathcal{A}_0$:
\begin{equation}
P_\mathcal{A}^\pm(\mathcal{A}_0) = \frac{-\dot{\mathcal{A}}^\pm(\mathcal{A}_0)}{32\pi^{3/2}\sqrt{\mathcal{A}_0}} = \frac{\mp \sqrt{M_S} }{4\pi^{3/4}\sqrt[4]{\mathcal{A}_0}}
\end{equation} 
This is because the momentum is zero only at infinity in this scenario. Note that the momentum switches sign when the branch changes, since (\ref{eqn:sdot_cla_pm2}) has opposite signs in the two branches.\\
The final solutions, after simplifying, read:
\begin{align}
&\epsilon = 0 \; : \qquad P_{\mathcal{A}}^\pm(\mathcal{A}) =\mp\sqrt[4]{\frac{M_S^2}{256\pi^3\mathcal{A}}}\\
&\epsilon = +1\; : \qquad P_{\mathcal{A}}^\pm(\mathcal{A}) = \mp\sqrt{\frac{\kappa_S\left(-1+\sqrt{\frac{\mathcal{A}_0}{\mathcal{A}}}\right)}{64 \pi^2  \Xi^2 }}
\end{align}
The $P_{\mathcal{A}}^+$ is the equation for the out-rising solution and $P_{\mathcal{A}}^-$ for the in-falling surfaces, and we will only keep the in-falling solution as before, since here the out-rising is unphysical for the same motivations. The signs are inverted since, from equation (\ref{eqn:A_clas_eom}),  the momentum has always the opposite sign of the velocity.\\
 Fig. \ref{fig:infalling_classical_q} shows the behaviour of the momentum against the surface. As it approaches the singularity it clearly diverges.
\\
Now we tackle the effective model.

\section{\label{sec:eff_OS}EFFECTIVE OS MODEL}
For this section our work will be centered on the construction of an effective dynamical model, where the effective dynamics is built to take into accounts the quantum effects rooted into the implementation of a polymer quantization of the theory\cite{Corichi_2007}.\\
First we will briefly resume what this quantization procedure implies, and then we will develop a new formulation for the OS model, where we apply quantum corrections to the classical dynamics and solve the corresponding equations for the sphere surface and momentum.\\
Non trivial effects will emerge in both the solutions, hinting at what a fully quantization of the model could imply.\\
Let's now introduce the polymer quantization scheme.
\subsection{\label{subsec:poly}Polymer Effective Theory}
\begin{figure}[ht]
	\centering
  \includegraphics[scale=0.65]{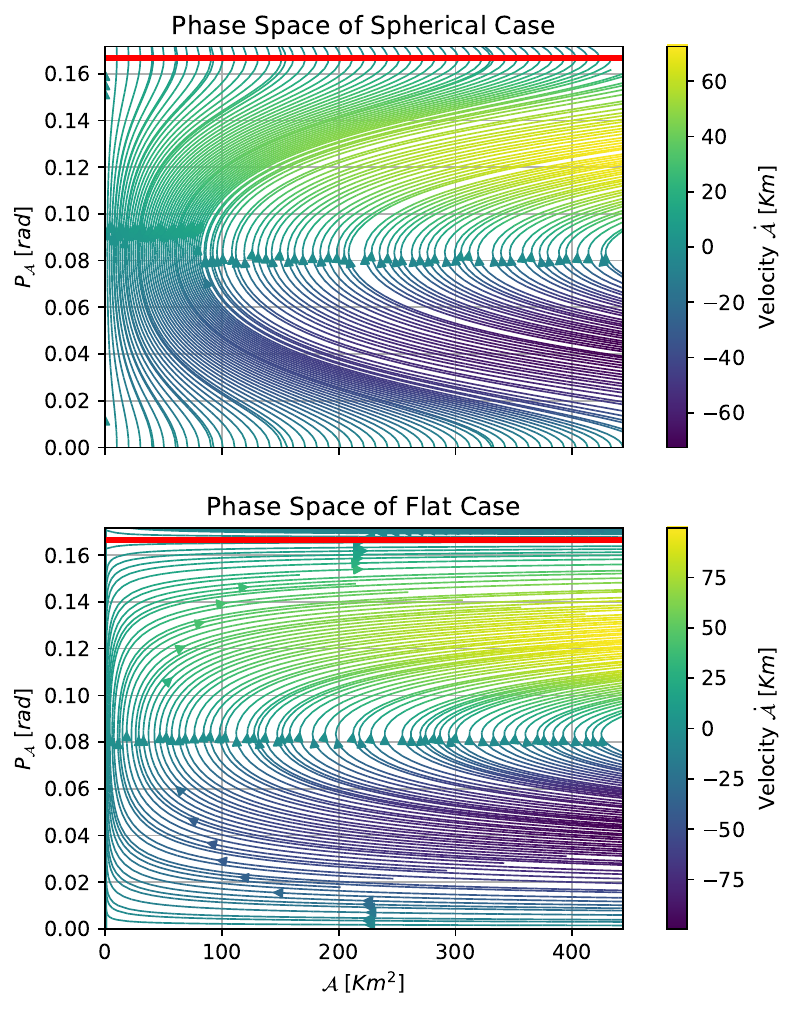}
  \caption{Reduced Phase Portraits in Geometrical Units for both the Flat and Spherical models. In red the line of constant $P_\mathcal{A} =p_0\pi$.}
  \label{fig:phase_red}
  \end{figure}

The framework of Polymer Quantization falls naturally into the quantization procedure referred as "Weyl Quantization", a generalization of the usual Heisenberg-Dirac procedure\cite{Strocchi_2016}.\\
In this framework the operators associated to the canonical variables $(q,p)$ are obtained by differentiating the corresponding one-parameter exponential operators, namely $U(\alpha) = e^{i\alpha q}$ and $V(\beta) = e^{i \beta p}$. When one of the exponential operators isn't continuous with respect to the parameter, the operator corresponding to its canonical variable is not defined, since the operation of derivation is ill defined. The polymer quantization is then the Weyl quantization associated to the invariance under the diffeomorphisms group\cite{Morchio_2007}, and contains multiple schemes.\\
We choose the polymer scheme in which the momentum operator in the momentum polarization is not defined, since this corresponds to a choice of discrete areas. This choice is inspired by the literature on loop quantum gravity\cite{Rovelli_1994}\cite{Rovelli_1995}.\\
Since the Hamiltonian in Eq. (\ref{eqn:ham_SQ}) contains the squared momentum $P_\mathcal{A}^2$, we need to regularise it by introducing a graph structure on the phase space, regularizing the $\mathcal{A}$ variable with a lattice of step $\mu_0$. \\
The introduction of this regularization implies that there is a cutoff on the momentum values, which we will define as $p_0 = \frac{\hbar}{\mu_0}$. To see how to implement such a structure into our effective model we show how it affects the quantum dynamics.\\
The operator action of the squared momentum is defined by:
\begin{equation}
    \hat{P}_\mathcal{A}^2 \phi(P_\mathcal{A}) = P_\mathcal{A}^2 \phi(P_\mathcal{A})
\end{equation}
Since we don't have such an operator, we approximate $P_\mathcal{A}$ with a translation operator $\hat{T}$:
\begin{equation}
    \hat{P}_\mathcal{A} \simeq \frac{\hbar}{\mu_0}\frac{1}{2i}\left(\hat{T}(\mu_0) -\hat{T}(-\mu_0)\right) 
\end{equation}
When this operator acts on wave functions we get:
\begin{equation}
    \hat{P}_\mathcal{A} \phi(P_\mathcal{A}) \simeq \frac{\hbar}{\mu_0}\frac{1}{2i}\left(e^{i\frac{\mu_0}{\hbar}P_\mathcal{A}}-e^{-i\frac{\mu_0}{\hbar}P_\mathcal{A}}\right) \phi(P_\mathcal{A})
\end{equation}
This means that we have the following relation:
\begin{equation}
    \hat{P}_\mathcal{A}^2 \phi(P_\mathcal{A}) \simeq p_0^2 \sin^2\left(\frac{P_\mathcal{A}}{p_0}\right)\phi(P_\mathcal{A})
\end{equation}
So we can obtain an effective model, with the natural inclusion of a momentum cutoff, by substituting the squared momentum with its corresponding polymer eigenvalue, namely by allowing $P_\mathcal{A}^2 \rightarrow p_0^2 \sin^2\left(\frac{P_\mathcal{A}}{p_0}\right)$.\\
After this substitution into the Hamiltonian, we can derive the effective equations of motion.\\
The effective Hamiltonian, in the comoving gauge, is given by:
\begin{equation}
    H_{pol} = -\frac{\Xi }{2}\left(32\pi^{\frac{3}{2}}p_0^2\sin^2\left(\frac{P_\mathcal{A}}{p_0}\right)\sqrt{\mathcal{A}} + \frac{\kappa_S}{\Xi ^2} \frac{\sqrt{\mathcal{A}}}{\sqrt{4\pi}}  -2 \frac{P_\tau }{\Xi }\right)
\end{equation}
The equations for the variables $P_\tau$ and $\tau$ are unchanged, so their solutions remain the same as equations (\ref{eqn:tau_tra}) and (\ref{eqn:Ptau_tra}). The equations for the dust momentum and area are instead modified by our quantum correction, and those read:
\begin{align}
			&\dot{\mathcal{A}} = \frac{\delta H_{pol}}{\delta P_\mathcal{A}} = -32 \pi ^{3/2}\Xi\; p_0\sin\left(\frac{P_\mathcal{A}}{p_0}\right)\cos\left(\frac{P_\mathcal{A}}{p_0}\right)\sqrt{\mathcal{A}}\label{eqn:HEQ_poly_V} \\
			&\dot{P}_\mathcal{A} = - \frac{\delta  H_{pol}}{\delta \mathcal{A}} =  \frac{\kappa_S}{8\Xi   \sqrt{\pi\mathcal{A}}} +\frac{8\Xi \pi ^{3/2} p_0^2\sin^2\left(\frac{P_\mathcal{A}}{p_0}\right)}{\sqrt{\mathcal{A}}}\label{eqn:HEQ_poly_P}
\end{align}
It is clear that now the momentum plays the role of an angle variable, with a period of $p_0\pi$.\\
This is also clear if we look at the phase space portraits.
\begin{figure}[ht]
	\centering
  \includegraphics[scale=0.55]{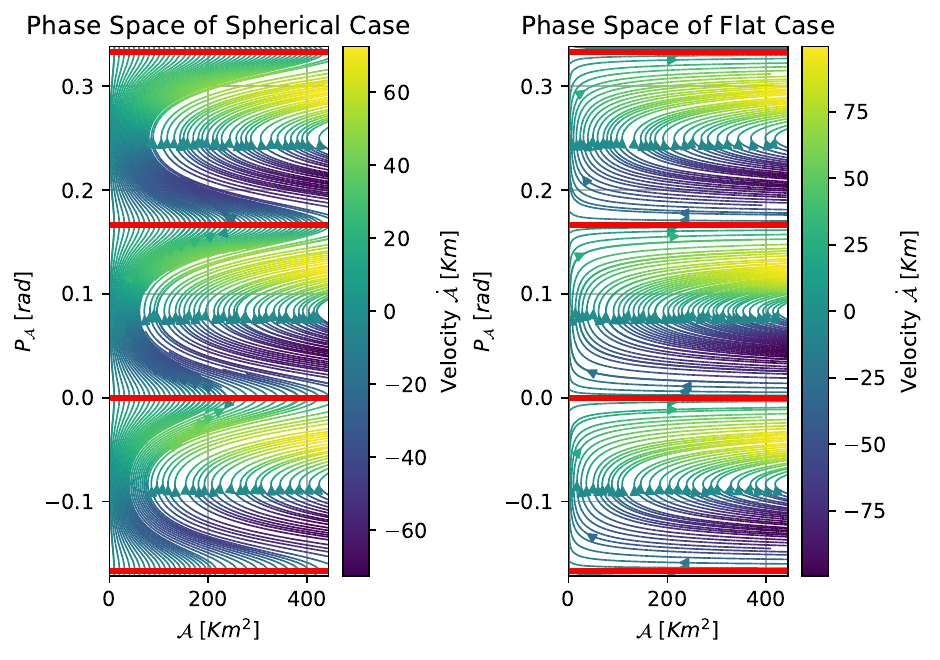}
  \caption{Phase Portraits in Geometrical Units for both the Flat and Spherical models. In red, lines of constant $P_\mathcal{A} =k p_0\pi$, $k\in \mathbb{Z}$.}
  \label{fig:phase_ext}
  \end{figure}
The phase space shown in Fig. \ref{fig:phase_ext} is clearly periodic, with the period previously given by $p_0\pi$. Being periodic, we can identify in the phase space $P_\mathcal{A} = P_\mathcal{A} + p_0\pi$, so that the momentum takes values on a circle. \\
This reduced phase space is shown in Fig. \ref{fig:phase_red}.
\\
It is immediately clear that the spherical case oscillates between a finite maximum and a finite minimum, while the flat case possesses a finite minimum but has its maximum at infinity. The analysis of those features is done in the subsections  \ref{subsec:eff_surface_eom} and \ref{subsec:eff_momentum_eom}.\\
We know seek the solutions of the effective equations (\ref{eqn:HEQ_poly_V}) and (\ref{eqn:HEQ_poly_P}).
\subsection{\label{subsec:eff_surface_eom}Sphere Surface Equation of Motion}
\begin{figure}[ht]
	\centering
  \includegraphics[scale=0.6]{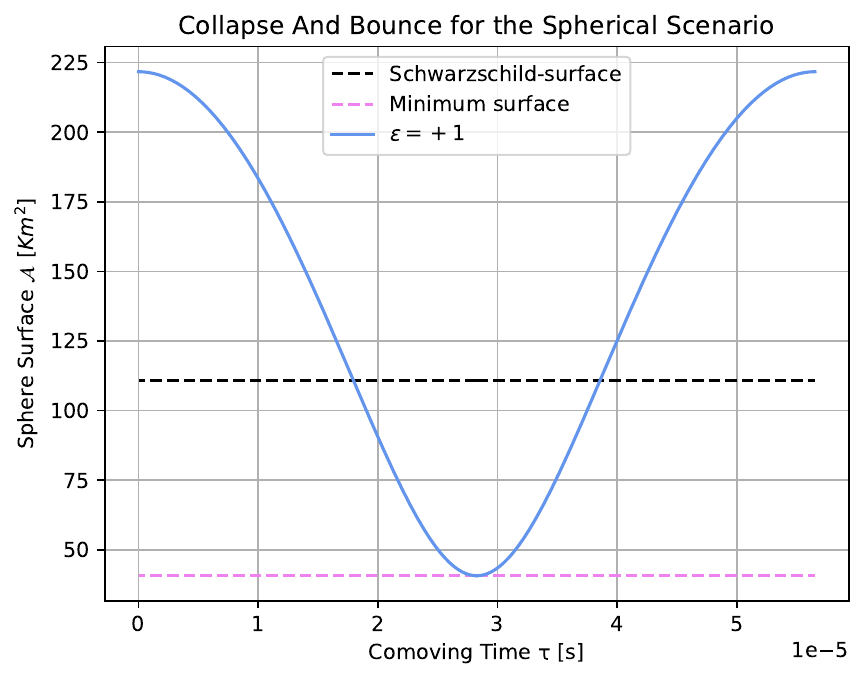}
  \caption{Bouncing object for the $\epsilon=+1$ scenarios. The parameters chosen are $\alpha = 2$ , $M_S = M_\odot$ and $\mu_0 = 6\pi\;\ell _p^2$\\The minimum is $\mathcal{A}_{min} = 40.6\;\text{Km}^2$.}
  \label{fig:poly_10_ep_1}
  \end{figure}

\begin{figure}[ht]
\centering
  \includegraphics[scale=0.6]{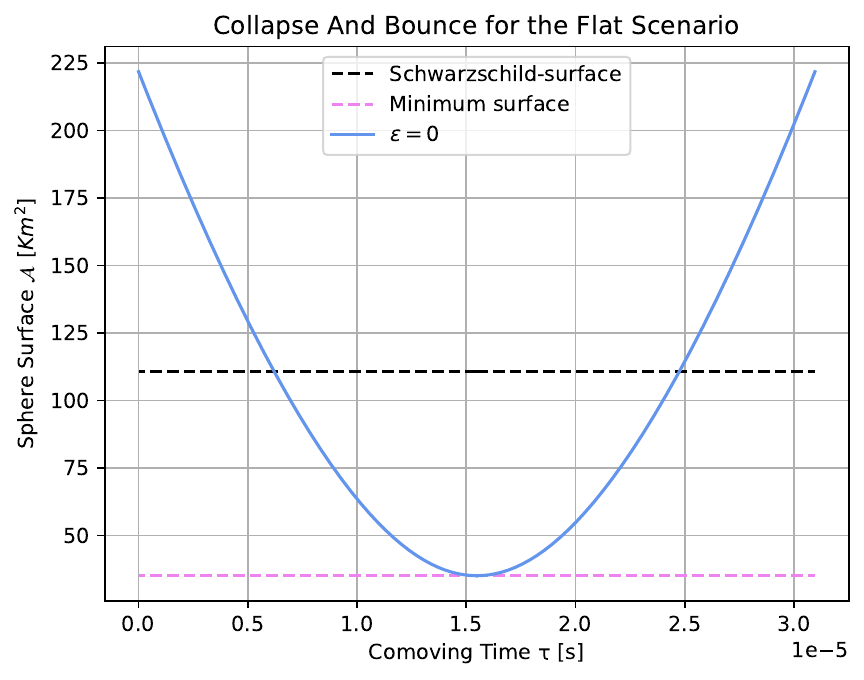}
  \caption{Bouncing object for the $\epsilon=0$ scenarios. The parameters chosen are $\alpha = 2$ , $M_S = M_\odot$ and $\mu_0 = 6\pi\;\ell_p^2$\\The minimum is $\mathcal{A}_m = 35.1\; \text{Km}^2$.}
  \label{fig:poly_10_ep_0}	
\end{figure}

We start by studying the equation of motion for the area variable. In order to do so, we adopt the same technique we used in the previous section to solve the classical equations of motion.\\
The first step is again to obtain $\sin^2\left(\frac{P_\mathcal{A}}{p_0}\right)$ as a function of $\mathcal{A}$ from the constraint, solving $\mathcal{H}_{poly} \approx 0$.\\
Defining:
\begin{equation}
\kappa_p = 64\pi^2\Xi^2p_0^2
\end{equation}
Such procedure gives:
\begin{equation}
\label{eqn:poly_constr}
\sin^2\left(\frac{P_\mathcal{A}}{p_0}\right)= \frac{4M_S\sqrt{\pi\mathcal{A}}  - \kappa_S \mathcal{A}}{\kappa_p\mathcal{A}}
\end{equation}
From this it is immediate to obtain $\cos^2\left(\frac{P_\mathcal{A}}{p_0}\right)$, which is just:
\begin{equation}
    \cos^2\left(\frac{P_\mathcal{A}}{p_0}\right) = 1 - \sin^2\left(\frac{P_\mathcal{A}}{p_0}\right)
\end{equation}
Setting for readability:
\begin{equation}
 \mathfrak{a} =  \frac{4M_S\sqrt{\pi\mathcal{A}}  - \kappa_S \mathcal{A}}{\kappa_p\mathcal{A}}
\end{equation}
We see that:
\begin{equation}
\sin\left(\frac{P_\mathcal{A}}{p_0}\right)\cos\left(\frac{P_\mathcal{A}}{p_0}\right) =\pm \sqrt{\mathfrak{a}}\sqrt{\mathfrak{1-a}}
\end{equation}
Substituting this into $\dot{\mathcal{A}}$, the equation of motion simply becomes:
\begin{equation}
\dot{\mathcal{A}}^\pm = \pm 32 \pi ^{3/2}\Xi\; p_0\sqrt{\mathcal{A}(\mathfrak{a}-\mathfrak{a}^2)}
\end{equation}
The plus and minus signs represent again the in-falling and out-rising solutions.\\
Making explicit $\mathfrak{a}$ yields:
\begin{equation}
\label{eqn:polyAEOM}
\begin{split}
    \dot{\mathcal{A}}^\pm = \pm 32 \pi ^{3/2}\Xi\; p_0 &\left[\frac{4 M_S\sqrt{\pi\mathcal{A}}  - \kappa_S  \mathcal{A}}{\kappa_p}\right.\\
    &\left. - \frac{\left(4 M_S\sqrt{\pi\mathcal{A}}  - \kappa_S \mathcal{A}\right)^2}{\kappa_p^2\mathcal{A}}\right]^{\frac{1}{2}}
\end{split}
\end{equation}
As before, instead of solving directly this couple of equations, we search for the maxima and minima for the dust ball surface.\\
We do this by setting $\dot{\mathcal{A}}^2=0$ and solving for $\mathcal{A}$ after the substitution of the constraint solution (\ref{eqn:poly_constr}). \\
The one for $\epsilon = +1$ the solutions are:
\begin{equation}
\label{eqn:maxmin_p1_poly}
\mathcal{A}_{max} = \frac{16\pi M_S^2}{\kappa_S^2} \qquad
\mathcal{A}_{min}= \frac{16 \pi M_S^2}{\left(\kappa_S +\kappa_p\right)^2}
\end{equation}
From the domain restrictions in equation (\ref{eqn:polyAEOM}) we see that $\mathcal{A}_{max}$ is a maximum, the same as in the classical case, while $\mathcal{A}_{min}$ is a minimum, now different from zero. The most important aspect is that on the minimum now our solution will not be singular anymore, allowing us to match the two branches through such point.\\
The solution for the $\epsilon=0$ case yields:
\begin{equation}
\label{eqn:maxmin_p0_poly}
\mathcal{A}_m = \frac{16\pi M_S^2}{\kappa_p^2}
\end{equation}
Which again, from the domain restrictions in equation (\ref{eqn:polyAEOM}), we see that $\mathcal{A}_m$ is a regular non-zero minimum, while the sphere surface has no maximum, just like in the classical case.\\
A more precise analysis of the physical meaning of such findings is given below in the sub-subsection \ref{subsubsec:minima_cond}.\\
With the maxima and minima in our hands we now turn to solve the equations of motion.\\
The equations (\ref{eqn:polyAEOM}) are separable, so the differential problem can be cast as:
\begin{equation}
\begin{split}
    &\frac{ d\mathcal{A}^\pm}{\sqrt{\left(4 M_S\sqrt{\pi } -\kappa_S  \sqrt{\mathcal{A}}\right) \left(\sqrt{\mathcal{A}} \left(\kappa_S +\kappa_p\right)-4M_S\sqrt{\pi }\right)}} \\
    &=\pm \frac{1}{2\Xi p_0\sqrt{\pi}} dt
\end{split}
\end{equation}
Setting $t=\tau$ from the comoving time solution, and making use of the integrals defined in Appendix \ref{app:integr}, we get the solutions:
\begin{equation}
\mathcal{A}^-: - \frac{1}{2\Xi p_0\sqrt{\pi}} \tau = C^- + \mathcal{I}^{EA}_\epsilon(\mathcal{A}) 
\end{equation}
\begin{equation}
\mathcal{A}^+: \frac{1}{2\Xi p_0\sqrt{\pi}} \tau = C^+ + \mathcal{I}_\epsilon^{EA}(\mathcal{A}) 
\end{equation}
 The integration constant for the in-falling surface is found just like before, by setting $\mathcal{A}(\tau=0) = \mathcal{A}_0$, yielding:
\begin{equation}
C^- = -\mathcal{I}^{EA}_\epsilon(\mathcal{A}_0) 
\end{equation}
An important note: since we can choose for the $\kappa_S=0$ case whatever initial surface we desire, it is important to choose one which is always bigger than the minimum. For the $\epsilon=+1$ case this is always assured by the fact that $\mathcal{A}_{max} > \mathcal{A}_{min}$ always.\\
Since now the minimum is not a singular point, we can ask for a continuity condition to find the constant $C^+$ inside the out-rising solution:
\begin{equation}
\mathcal{A}^-(\tau=\tau_{min}) = \mathcal{A}^+(\tau=\tau_{min}) 
\end{equation}
Where $\tau_{min}$ is given by:
\begin{equation}
\tau_{min} =-2\Xi p_0\sqrt{\pi}\left[-\mathcal{I}^{EA}_\epsilon(\mathcal{A}_0) +\mathcal{I}^{EA}_\epsilon(\mathcal{A}_{min}) \right]
\end{equation}
This gives us:
\begin{equation}
C^+ = + \mathcal{I}^{EA}_\epsilon(\mathcal{A}_0) -2 \mathcal{I}^{EA}_\epsilon(\mathcal{A}_{min})
\end{equation}
By substitution in the above integrals of the respective minima, it is easy to see that $\mathcal{I}^{EA}_\epsilon(\mathcal{A}_{min}) = 0$ for both cases. \\
The constant for the out-rising solution then become:
\begin{equation}
C^+ = + \mathcal{I}^{EA}_\epsilon(\mathcal{A}_0)
\end{equation}
The meaning of such an integration condition lies in the fact that $\tau$, the comoving time, is continuous as a function of $\mathcal{A}$. This implies that there is no reason to restart the clock once the dust ball has reached the minimum.\\
Finally the equations of motion read:
\begin{align}
\mathcal{A}^-&: - \frac{1}{2\Xi p_0\sqrt{\pi}} \tau = -\mathcal{I}^{EA}_\epsilon(\mathcal{A}_0) + \mathcal{I}^{EA}_\epsilon(\mathcal{A})\label{eqn:inf-pol} \\
\mathcal{A}^+&: \frac{1}{2\Xi p_0\sqrt{\pi}} \tau =+ \mathcal{I}^{EA}_\epsilon(\mathcal{A}_0) + \mathcal{I}^{EA}_\epsilon(\mathcal{A})\label{eqn:ouris-pol} 
\end{align}
A plot of the trajectory, against comoving time, is shown in Fig. \ref{fig:poly_10_ep_1} and Fig. \ref{fig:poly_10_ep_0} for the two cases.
  \\
The trajectories show clearly that the dust ball bounces, so that after the minimum it starts to expand again, eventually crossing the event horizon.\\
In the regime where the quantum effects become important, namely the high momentum regime, a change in the geometry is induced by quantum effects, and a black hole geometry makes a transition to a white hole geometry. This will be analyzed clearly in the sub-subsection \ref{subsubsec:transit}.\\
Now let's review the physical meaning of the minima we have found.
\subsubsection{\label{subsubsec:minima_cond}Minima and Cut-Offs}
Now we analyze more critically the significance of the minima we have found.\\
The graph structure on the phase space that we introduced was meant to regularize the $P_\mathcal{A}^2$ operator, since this is not a sensible operator in the full quantum theory.
The quantum theory suggests that only finite area shift operators exist, and that the infinitesimal version, the $P_\mathcal{A}$ operator, is just a classical approximation which emerges for low momenta. This means that a full quantum theory could possess some sort of minimal area gap, which is small but not infinitesimal, which we emulated in our effective theory with the parameter $\mu_0$\cite{Ashtekar_2002}.\\
The minima that we have found in this work are much bigger than simply $\mu_0$, in fact they also strictly depend on the physical parameters of the model and on the specific Hamiltonian which describe the collapse. What this means is that our newly found minima arise dynamically and are not the fundamental area gap $\mu_0$.\\
The presence of such dynamical minima is nonetheless related to the existence of a fundamental discreteness of the area, which enables some kind of quantum gravitational pressure to slow the collapse and halt it at a non zero value of the surface, ultimately creating a bouncing object. Like  in the case of the neutron stars or white dwarfs, a quantum pressure has macroscopical effects on the collapse.\\
The minima clearly go to zero when we take the limit $\mu_0 \rightarrow 0$, equivalent to the $p_0\rightarrow\infty$ limit, so $\mu_0$ is supposed to be the source of such quantum gravitational pressure. In subsection \ref{subsec:pressure} we will characterize such phenomenon more accurately.\\
We also want to notice that the points $\mathcal{A}_{min}$ and $\mathcal{A}_{m}$ could lie outside the Schwarzschild-surface for some values of the minimum area gap.\\
Before analyzing such conditions, we recall that $\hbar$ is expressed in geometrized units and it corresponds to $\hbar = \ell_p^2$, so $p_0 = \ell_p^2/\mu_0$ is the rateo between the Planck area and the minimum area gap.\\
We also define $\mathcal{A}_p = 4\pi \ell_p^2 = 4\pi\hbar$ to be the Planck area of a sphere of Planck radius.\\
We now see for what values of the parameter $\mu_0$ we have a dynamical minimum bigger than the Schwarzschild-surface.\\ 
In the $\epsilon=0$ case the condition $\mathcal{A}_m > \mathcal{A}_{SC}$ for a minimum outside the event horizon holds only if:
\begin{equation}
\frac{16\pi M_S^2}{\kappa_p^2}>16 \pi  M_S^2 \Longrightarrow \mu_0 >8 \pi \ell_p^2 
\end{equation}
For the $\epsilon=+1$ case the condition  $\mathcal{A}_{min} > \mathcal{A}_{SC}$  holds only if:
\begin{equation}
\frac{16 \pi M_S^2}{\left(\kappa_S+\kappa_p\right)^2} > 16 \pi M_S^2 \Longrightarrow  \frac{\ell_p^2}{\mu_0}<\frac{\sqrt{\frac{1-\kappa_S}{\Xi ^2}}}{8 \pi }
\end{equation}
So, after a quick substitution, the conditions to have a dynamical minimum outside the event horizon, in both cases, are given by:
\begin{equation}
\label{eqn:min_outside}
\begin{split}
\epsilon = 0:& \qquad \mu_0 > 2\mathcal{A}_p\\
\epsilon = +1:& \qquad \mu_0 >  2\mathcal{A}_p \frac{\Xi}{\sqrt{1-\kappa_S}}
\end{split}
\end{equation}
Those are reasonable parameters for a quantum cut-off, being of the same order of the Planck-sphere area, and those may have a macroscopic effect on the dust ball collapse, since for such values the formation of an event horizon is avoided.\\
Such conditions could then pose a constraint on the minimum area gap.\\
We now solve the equation of motion for the momentum.
\subsection{\label{subsec:eff_momentum_eom}Sphere Momentum Equation of Motion}
\begin{figure}[ht]
	\centering
  \includegraphics[scale=0.6]{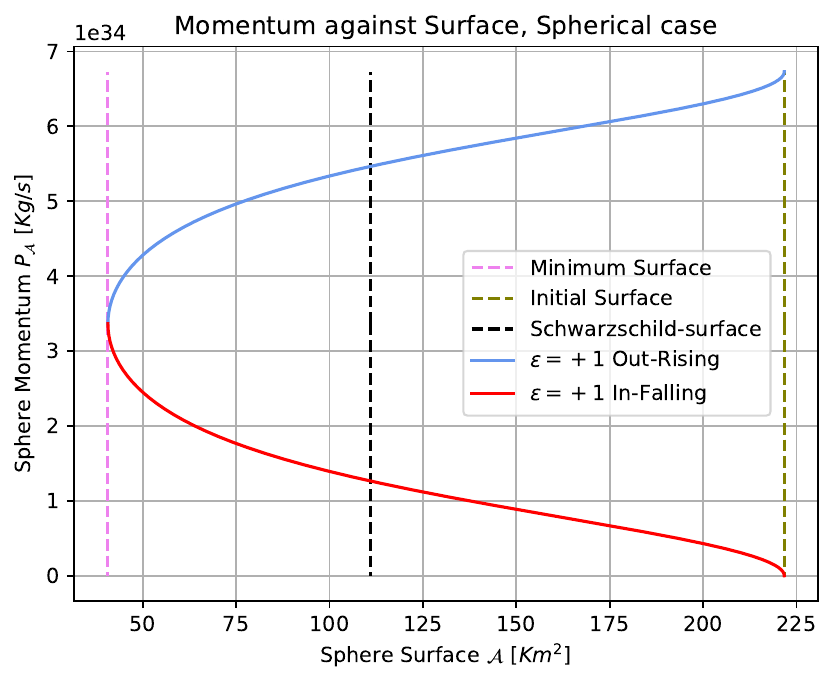}
  \caption{Momentum as a function of the surface for the $\epsilon=+1$ scenario. The parameters chosen are $\alpha = 2$ , $M_S = M_\odot$ and $\mu_0 =6\pi \ell_p^2$}
  \label{fig:poly_mom_p1}
\end{figure}

\begin{figure}[ht]
	\centering
  \includegraphics[scale=0.6]{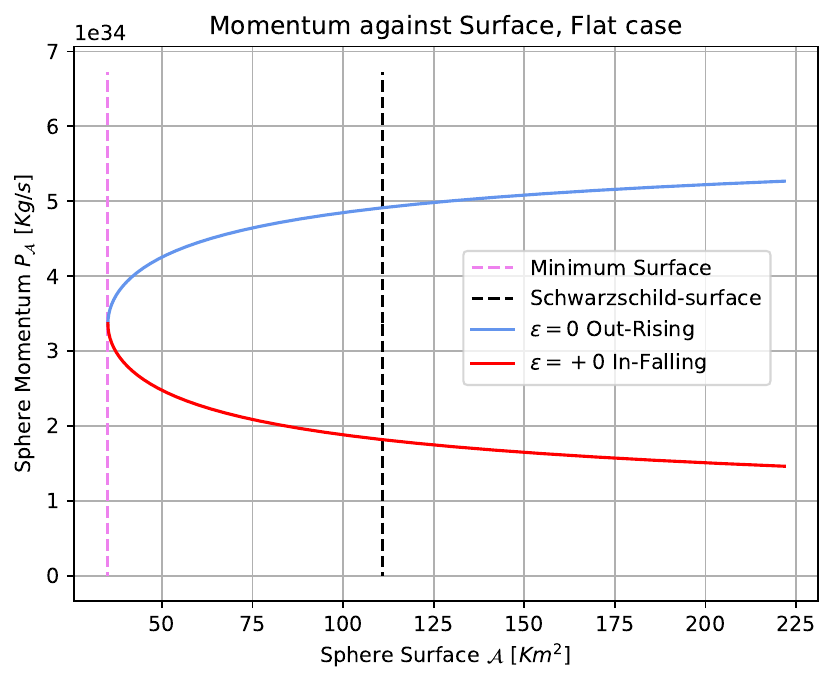}
  \caption{Momentum as a function of the surface for the $\epsilon=0$ scenario. The parameters chosen are $\alpha = 2$ , $M_S = M_\odot$ and $\mu_0 = 6\pi \ell_p^2$}
  \label{fig:poly_mom_p0}
\end{figure}

As in the classical section, we will not solve the equation in function of the comoving time $\tau$ but again we aim to solve for $P_{\mathcal{A}}$ as a function of $\mathcal{A}$.\\
We recall that $P_{\mathcal{A}}'= \frac{dP_{\mathcal{A}}}{d\mathcal{A}} =\frac{\dot{P}_{\mathcal{A}}}{\dot{\mathcal{A}}}$, so we substitute the constraint solution (\ref{eqn:poly_constr}) inside $\dot{\mathcal{A}}$ and $\dot{P}_{\mathcal{A}}$, thus obtaining the equation:
\begin{equation}
P_{\mathcal{A}}' = \pm \frac{P_\tau }{64 \pi ^{3/2}\Xi\; p_0\mathcal{A}\sqrt{\frac{4 M_S\sqrt{\pi\mathcal{A}}  -\kappa_S\mathcal{A}}{\kappa_p} - \frac{\left(4 M_S\sqrt{\pi\mathcal{A}}  - \kappa_S \mathcal{A}\right)^2}{\kappa_p^2\mathcal{A}}}}
\end{equation}
The plus and minus equations are associated to the in-falling and out-rising velocity equations.\\
We can directly integrate those two equations without the need of defining some external integrals and without treating separately the flat and spherical case, since if we substitute $\kappa_S\rightarrow0$ and $\Xi\rightarrow1$ in the solutions we obtain the same expression as if we evaluate the two integrals separately.\\
The solution then is:
\begin{equation}
\begin{split}
P_{\mathcal{A}}^\pm(\mathcal{A}) = &\pm p_0 \text{arctan}\left(\sqrt{\frac{-4 M_S\sqrt{\pi } + \sqrt{\mathcal{A}} \left(\kappa_S+\kappa_p\right)}{4 M_S\sqrt{\pi }  -\kappa_S \sqrt{\mathcal{A}}}}\right) \\
&+C_\pm
\end{split}
\end{equation}
To find the integration constant $C_-$ we set the momentum to be zero at the initial surface $\mathcal{A}_0$, which is $\mathcal{A}_{max}$ for the spherical case and $\mathcal{A}=+\infty$ for the flat case. In both the cases the argument of the arctangent goes to infinity.\\
This yields:
\begin{equation}
  0 = - p_0\frac{\pi}{2} + C_- \Longrightarrow C_- = p_0\frac{\pi}{2}
\end{equation}
We now need to impose the continuity on the minimum for the out-rising branch:
\begin{equation}
   P_{\mathcal{A}}^-(\mathcal{A}_{min}) =   P_{\mathcal{A}}^+(\mathcal{A}_{min})
\end{equation}
The arctangent is zero on the minimum, so we obtain:
\begin{equation}
    p_0\frac{\pi}{2} =   C_+
\end{equation}
So the two constants are $C_- = p_0\frac{\pi}{2} = C_+$ for both flat and spherical cases, and the solution reads:
\begin{align}
    P_{\mathcal{A}}^- &= -p_0\left[ \text{arctan}\left(\sqrt{\frac{-4M_S\sqrt{\pi } + \sqrt{\mathcal{A}} \left(\kappa_S+\kappa_p\right)}{4 M_S\sqrt{\pi }  -\kappa_S \sqrt{\mathcal{A}}}}\right) -\frac{\pi}{2}\right] \\
    P_{\mathcal{A}}^+ &= + p_0 \left[ \text{arctan}\left(\sqrt{\frac{-4M_S\sqrt{\pi } + \sqrt{\mathcal{A}} \left(\kappa_S+\kappa_p\right)}{4 M_S\sqrt{\pi }  -\kappa_S \sqrt{\mathcal{A}}}}\right)  + \frac{\pi}{2}\right]
\end{align}
Where $P_{\mathcal{A}}^-$ is the in-falling momentum and $P_{\mathcal{A}}^+$ is the out-rising one.\\
We show in Fig. \ref{fig:poly_mom_p1} and Fig. \ref{fig:poly_mom_p0} the trajectories for the momentum in both the spherical and flat case. \\
As can be seen, in the spherical case the sphere has zero momentum at surface $\mathcal{A}_0$, and the same is true for the flat case, which has it at $\mathcal{A}=\infty$, so at finite surface $\mathcal{A}_0$ it possesses a non-zero momentum.\\
Since the arctangent is zero on the minima,  the momentum value $P_\mathcal{A}(\mathcal{A}_{min}) = p_0\frac{\pi}{2}$ is the one associated to the velocity inversion.
\\
We also notice that while in equation (\ref{eqn:A_clas_eom}) there is a minus sign, so that the momentum needs always to have the opposite sign of the velocity, in the effective case this is not true, due to the velocity inversion after reaching the minimum.\\
We will tackle the meaning of such inversion in subsubsection \ref{subsubsec:transit}.
\\
We now aim to analyze the inversion behaviour.
\subsubsection{\label{subsubsec:transit} Velocity Inversion and Transition Amplitudes}
\begin{figure}[ht]
	\centering
  \includegraphics[scale=0.45]{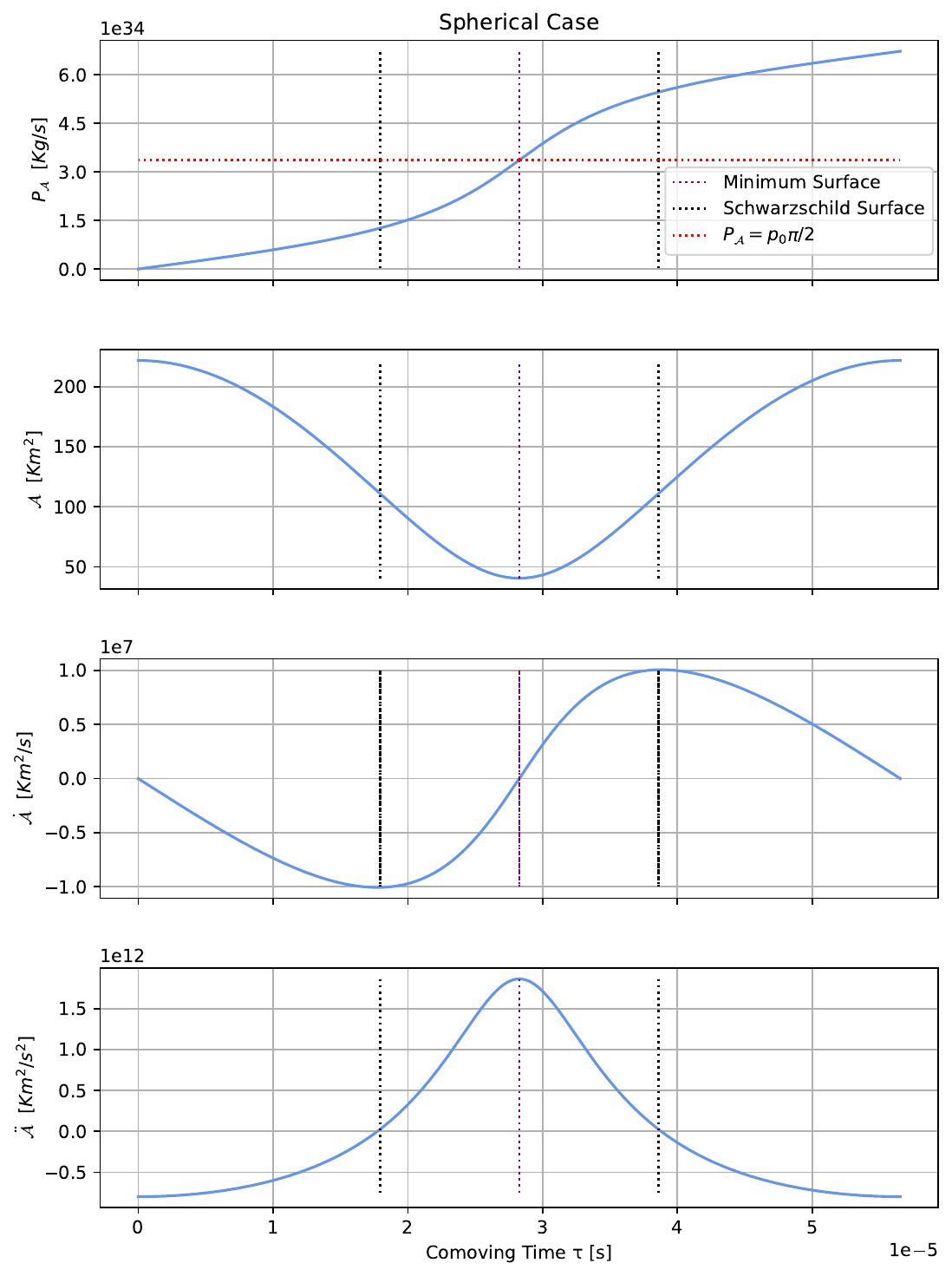}
  \caption{Time evolution of the system in the $\epsilon=+1$ scenario. The parameters chosen are $\alpha = 2$ , $M_S = M_\odot$ and $\mu_0 =6\pi \ell_p^2$.\\The vertical lines show the time when the surface reaches the minimum, the inflection point and the Schwarzschild surface.}
  \label{fig:sph}
\end{figure}

\begin{figure}[ht]
	\centering
  \includegraphics[scale=0.45]{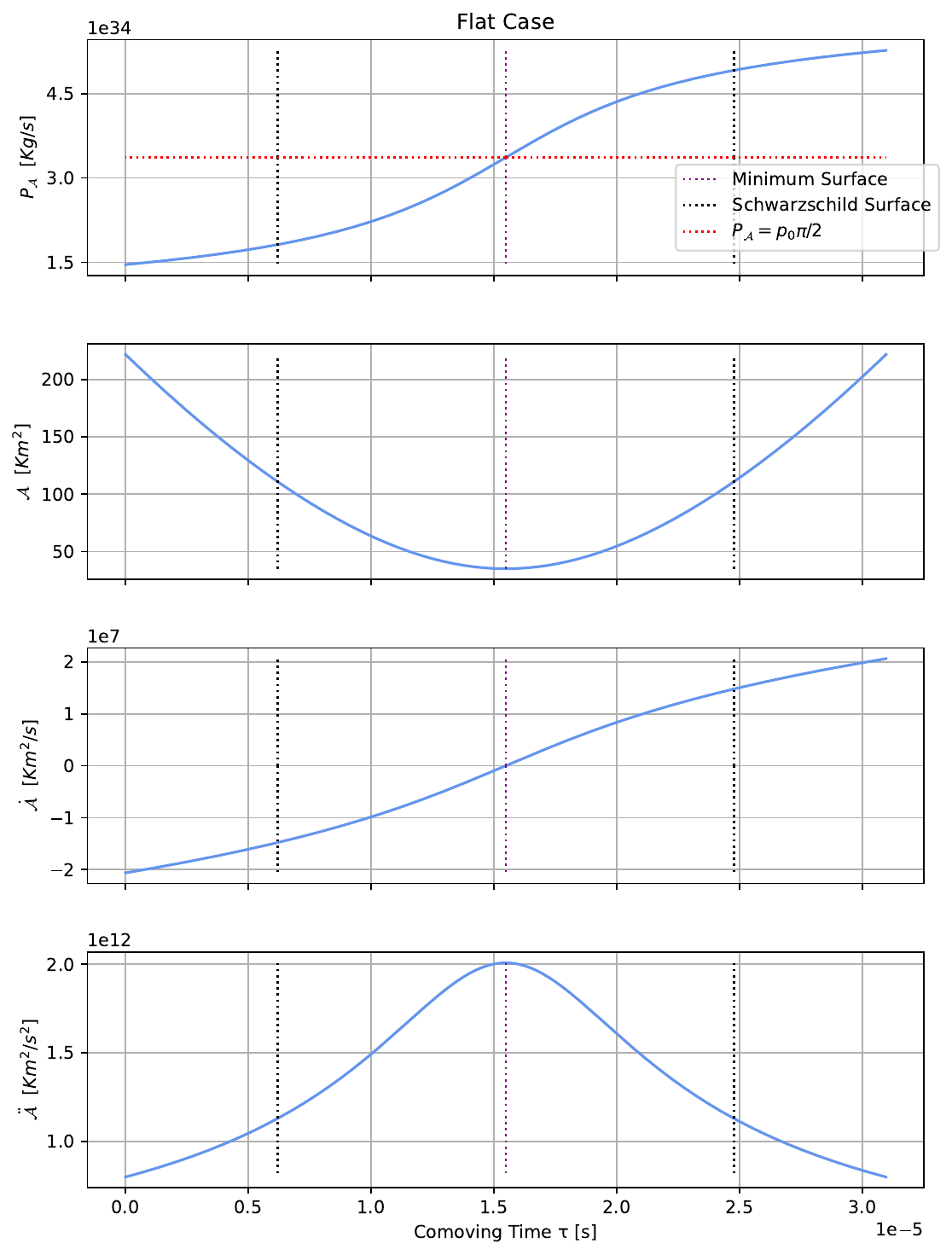}
  \caption{Time evolution of the system in the $\epsilon=0$ scenario. The parameters chosen are $\alpha = 2$ , $M_S = M_\odot$ and $\mu_0 =6\pi \ell_p^2$.\\The vertical lines show the time when the surface reaches the minimum and the Schwarzschild surface.}
  \label{fig:flt}
\end{figure}

We are ready to understand the implications of the inversion of the velocity.\\
To understand the physical meaning behind this behaviour, we will firstly analyze the time evolution of the system, to check whether it is continuous or not. \\
The first thing we need to do is to evaluate $\ddot{\mathcal{A}}$. Using the fact that:
\begin{equation}
\label{eqn:def_acc}
\ddot{\mathcal{A}} = \frac{d\dot{\mathcal{A}}}{dt} = \frac{\partial \dot{\mathcal{A}}}{\partial t} + \left\lbrace \dot{\mathcal{A}}, H\right\rbrace =  \dot{\mathcal{A}}\frac{\partial \dot{\mathcal{A}}}{\partial \mathcal{A}} + \dot{P}_\mathcal{A} \frac{\partial \dot{\mathcal{A}}}{\partial P_\mathcal{A}}
\end{equation}
We can obtain the surface acceleration:
\begin{equation}
\label{eqn:poly_acc}
\ddot{\mathcal{A}}  =-4 \pi\kappa_S + 8 \pi  \left(\kappa_S+\kappa_p/2\right) \sin ^2\left(\frac{ P_\mathcal{A}}{p_0}\right)
\end{equation}
We now show in Fig. \ref{fig:sph} and Fig. \ref{fig:flt} the trajectories against time of $\mathcal{A},\;\dot{\mathcal{A}}, \;\ddot{\mathcal{A}}$ and $P_\mathcal{A}$. The acceleration in the spherical case has an inflection point:
\begin{equation}
\mathcal{A}_{IN} = \mathcal{A}_{max}\frac{ \left(\kappa_S+\kappa_p/2\right)^2}{\left(\kappa_S+\kappa_p\right)^2}
\end{equation}
In the flat case there are no inflection points. In the flat case is also clearly visible that both the momentum and the velocity are not zero on some surface $\mathcal{A}_0$ at a finite time, except for the zero velocity on the minimum: that is compatible with the fact that the zero velocity condition is imposed at infinity.\\ 
 \\
The inversion has to be treated differently depending on whether it happens inside or outside the event horizon. When it happens outside the event horizon, there are no problems, since it means that the dust particles 'bounce' at a finite radius, like in the presence of a potential barrier, and turn away from the event horizon. \\
The problem appears if such an inversion happens after entering the event horizon: inside the  Schwarzschild surface the velocity can only point in the decreasing radial direction, so the meaning of the inversion is that of an inversion of causality, namely a transition to a white hole geometry with an anti-horizon.\\
This could happen due to a quantum effect yet to be explored, and that make possible the transition between the two geometries: that is what in the literature is known as a black-to-white hole geometry transition\cite{Haggard2015}\cite{Mnch2021}\cite{Kelly2020}\cite{Bianchi2018}\cite{Achour2020}.\\
This transition should happen when the corresponding location in the phase space for the canonical variables is in a neighborhood of the point $(\mathcal{A}_{min},\;p_0\frac{\pi}{2})$, associated to the minimal surface and inversion value of the momentum, with $\mathcal{A}_{min} < \mathcal{A}_{SC}$. The inversion value of the momentum is clearly Planckian, being of the same order of $M_p / t_p$.\\
This is where we expect the quantum effects to have a strong influence on the dynamics of the system.\\
We also suppose that the bounce could be altered in a more accurate model. \\
This is because the dust model neglects all the internal thermodynamical properties, as well as all the internal degrees of deformation and compressibility. Those could result, in a more realistic model, in a different transition mechanism between the two regions, where the momentum interacts with the internal degrees of freedom of the star to over-compress it and change its thermodynamical properties. This, together with the inclusion of dissipative effects, could avoid the bounce once inside the event horizon and stabilize the sphere at some fixed surface less than the Schwarzschild surface, but still not singular. \\
Alternative bounce models with and without dissipative effects can be found at \cite{Rovelli_2018}\cite{DeLorenzo2016}\cite{Husain2022}\cite{Achour2020v1}\cite{BenAchour2020}\cite{Barca2023}.\\
Also in string models the singularity is avoided due to T-duality effects \cite{Jusufi2023}.\\ 
Inhomogeneous density profiles show a very interesting property, that is, the formation of shock-waves due to the most inner shells, which bounce first, colliding with the outer layers of the cloud which are still collapsing. As a result, the horizon disappear when the shock reaches said surface.\\
Those effects are not present is the OS model, being homogeneous.\\
The presence of such phenomena in an LQG setting was studied extensively in \cite{Fazzini2023} \cite{Fazzini2024} \cite{HusainRev2022}.\\
We expect similar effects to take place also in the polymer model, when pressure and inhomogeneities are introduced, and those should alter significantly the bounce dynamics.\\
We are ready to analyze some phenomenological aspects of this effective model.
\section{\label{sec:phen}PHENOMENOLOGY}
The scope of this section is to analyze the implications of the newly found model.\\
First we are gonna characterize how the polymer quantization effects could be understood as a form of quantum gravitational pressure.\\
Then we will analyze the implications, divided in two categories: the phenomena which arise already at a macroscopical scale, before the sphere reaches the quantum regime, and the effects that take place only after the system has entered the quantum region. Both regimes could have observational consequences.
\subsection{\label{subsec:pressure}Polymer Pressure}
Before diving into the description of possibly observable effects of this model, we wish to give a proper description of the polymer quantum pressure mentioned to be the microscopic cause of the modified collapse.\\
A similar quantum effective pressure was also found in \cite{Bambi2013} and \cite{Liu2014}.\\
Clearly there is no actual macroscopical pressure inside the dust, but as we will see, the effective dynamics generated by the polymer implementation of the area gap can be thought as the introduction of an internal pressure generated by quantum effects of the gravitational field itself.\\
To achieve this result, we first recall the Friedmann acceleration equation for a homogeneous isotropic fluid, which reads:
\begin{equation}
\frac{\ddot{a}}{a} = -\frac{4\pi}{3}\left(\rho + 3p\right)
\end{equation}
We can rewrite this equation as:
\begin{equation}
\frac{\ddot{R}_s}{R_s} = -\frac{4\pi}{3}\rho -4\pi p
\end{equation}
Since the LHS is given by:
\begin{equation}
\label{eqn:fried_acc}
\frac{\ddot{R}_s}{R_s} = \frac{\ddot{\mathcal{A}}}{2\mathcal{A}} - \frac{\dot{\mathcal{A}}^2}{4\mathcal{A}^2}
\end{equation}
This can be evaluated using equation (\ref{eqn:def_acc}) for the classical case, and yields:
\begin{equation}
\left.\frac{\ddot{R}_s}{R_s}\right|_{CL} = -\frac{8\pi^{3/2}M_S}{\mathcal{A}^{3/2}} = -\frac{4\pi}{3}\rho
\end{equation}
So there is no pressure in the classical case, as expected. For the polymer case the situation is different.\\
Using equation (\ref{eqn:poly_acc}) we see that:
\begin{equation}
\left.\frac{\ddot{R}_s}{R_s}\right|_{POL} = -\frac{8\pi^{3/2}M_S}{\mathcal{A}^{3/2}} +\frac{M_S^2}{\mathcal{A}^2\Xi^2 p_0^2} -\frac{\left(\frac{\kappa_SM_S \sqrt{\mathcal{A}}}{\sqrt{\pi }}\right)}{4 \mathcal{A}^2 p_0^2}
\end{equation}
So there are extra terms, coming from the introduction of the polymer corrections, which we will call polymer pressure:
\begin{equation}
-4\pi \mathcal{P}_{\mu_0} = +\frac{M_S^2}{\mathcal{A}^2\Xi^2 p_0^2} -\frac{\left(\frac{\kappa_SM_S \sqrt{\mathcal{A}}}{\sqrt{\pi }}\right)}{4 \mathcal{A}^2 p_0^2}
\end{equation}
Rewriting this in a more clean way gives:
\begin{equation}
\label{eqn:poly_press}
 \mathcal{P}_{\mu_0}  = \frac{\mathcal{A}_{SC}}{\kappa_p} \frac{1}{\mathcal{A}^2}\left(\sqrt{\frac{\mathcal{A}}{\mathcal{A}_0}}-1\right)
\end{equation}
Now let's analyze the pressure $\mathcal{P}_{\mu_0}$.\\
The quantum origin is very clear: if we take $\mu_0 = 0$ then $\frac{1}{\kappa_p} = 0$, so the pressure vanishes identically. Then we see that the pressure vanishes also at the start of the collapse, when $\mathcal{A}=\mathcal{A}_0$. This is good since we have seen that quantum effects arise with the increasing of the momentum.\\
The last thing to notice is that since $\mathcal{A}<\mathcal{A}_0$ always, the pressure is always negative: this means that $\mathcal{P}_{\mu_0}$ has to be interpreted as a kind of 3D tension, which tries to restore the surface to its initial value when it gets compressed.\\
An heuristic tentative explanation for that phenomenon is that the quanta of surface make resistance when compressed against each other, due to gravitational self interactions. In the classical scenario those quanta are not present and hence this new pressure is absent.\\
For this reasons from now on we will refer to this quantum gravitational pressure as "polymer tension".\\
With the nature of this tension made clear, we are now ready to study the phenomenological aspects of this model.
\subsection{\label{subsec:macro}Macroscopic Phenomena}
The first thing we wish to accomplish is to understand whether we need to impose some consistencies conditions on the newly introduced parameter $\mu_0$.\\
To do such a thing we need to check if the physical motion of the sphere exceeds the light speed limit. Since we are working with a ball of pressureless homogeneous dust, the velocity of the dust particles corresponds to the radial velocity of the star.\\
We then proceed to check if this velocity is consistent with the light speed limit. The radial velocity must be obtained from the surface velocity, since after the introduction of the effective correction we have lost the ability to make canonical transformations from $P_\mathcal{A}$ to $P_s$, but we can still obtain $R_s$ from $\mathcal{A}$.\\
We then have:
\begin{equation}
    \dot{\mathcal{A}} = \frac{d\mathcal{A}}{dR_s}\frac{dR_s}{dt} = \frac{d\mathcal{A}}{dR_s} \dot{R}_s
\end{equation}
Since $ \frac{d\mathcal{A}}{dR_s} = 8\pi R_s$, after the substitution of $\mathcal{A}(R_s)$, we get:
\begin{equation}
    \dot{R}_s = \frac{\dot{\mathcal{A}}}{8\pi R_s} = -8\pi\Xi p_0 \sin\left(\frac{P_\mathcal{A}}{p_0}\right)\cos\left(\frac{P_\mathcal{A}}{p_0}\right)
\end{equation}
We then make the trigonometric substitution $2\sin(x)\cos(x) = \sin(2x)$, so we end up with the equation of motion:
\begin{equation}
\label{eqn:radvel}
    \dot{R}_s = -4\pi\Xi p_0 \sin\left(\frac{2P_\mathcal{A}}{p_0}\right)
\end{equation}
Clearly the velocity is dependent on the value of the parameter $\mu_0$, so we need to impose by hand the subluminal collapse speed condition. It is easy to see that the velocity has its maximum value when the sine is equal to one, or minus one equivalently.\\
This means that the absolute maximum value of the velocity is:
\begin{equation}
    \dot{R}_s^{MAX} = 4\pi\Xi p_0 
\end{equation}
To impose $\dot{R}_s^{MAX}$ to be subluminal, in geometric units, means that we need $\dot{R}_s^{MAX} < 1$, yielding the following constraint on the parameter $\mu_0$:
\begin{equation}
\label{eqn:poly_subl}
    \frac{4\pi\Xi \ell_p^2}{\mu_0} < 1 \Longrightarrow \mu_0 > \mathcal{A}_p \Xi
\end{equation}
This means that we have a lower constraint on the polymer area gap given by $\mu_0 > \mathcal{A}_p \Xi$. Remember that in the flat case $\Xi = 1$.\\
Equivalently the constraint reads:
\begin{equation}
    p_0 < \frac{1}{4\pi\Xi}
\end{equation}
With those constraints established, we see that the values for which we have a minimum outside the event horizon in equations (\ref{eqn:min_outside}) are compatible with what we have found.\\
Regarding only the spherical case, which is a model more physically relevant than the flat one, we can investigate deeper into the consequences of such scenario.
\subsubsection{\label{subsubsec:new_obj}A New Astrophysical Object}
\begin{figure}[ht]
	\centering
  \includegraphics[scale=0.6]{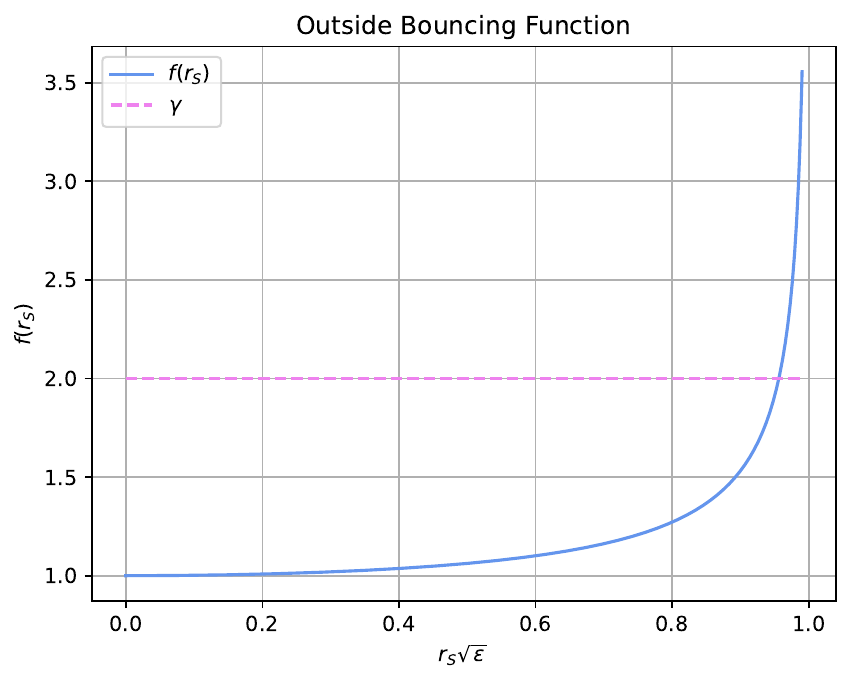}
  \caption{Behaviour of $f(r_S)$ as a function of $r_S$. The function clearly diverges as $r_S$ approaches 1. \\$\gamma$ is set to 2 for illustrative reasons.}
  \label{fig:gamma_sphere}
  \end{figure}

Let us fix a value of the parameter $\frac{\mu_0}{\mathcal{A}_p} = \gamma$. Then the condition for a minimum outside the event horizon reads:
\begin{equation}
    \frac{2\Xi}{\sqrt{1-\kappa_S}} < \gamma
\end{equation}
To see when this condition is satisfied we explicit the $r_S$ parameter in the LHS, which yields the function:
\begin{equation}
    f(\kappa_S)=\frac{4 \kappa_S ^{3/2}}{3 \sqrt{1-\kappa_S} \left(2 \text{arctan}\left(\frac{-1+\sqrt{\kappa_S}}{\sqrt{1-\kappa_S}}\right)+\pi - \sqrt{\kappa_S -\kappa_S^2 }\right)}
\end{equation}
Recall that the $r_S$ dependence is recovered by $\kappa_S$ since $\sqrt{\kappa_S} = r_S\sqrt{\epsilon}$, where $\epsilon=1$ is kept for dimensional reasons.\\
We show in Fig. \ref{fig:gamma_sphere} the behaviour of such function with $r_S \in (0,1)$, as imposed by the condition (\ref{eqn:epsparam}).\\
\\
The plot clearly shows that the closer to the event horizon the collapse start, i.e. the closer we get to $\kappa_S = 1$, the greater the value of the area gap must be in order to avoid the formation of the event horizon.\\
The plot gives the values of $\kappa_S$, far enough from the event horizon given a value of $\gamma$, for which the sphere will oscillate between the initial collapsing size $\mathcal{A}_0$ and the minimum value $\mathcal{A}_{min}$, outside its event horizon.\\
This suggests that some astrophysical objects, massive enough that should form a black hole instead of a neutron star, could instead become some new radially pulsating objects, given the right conditions on the size and the polymer parameter.\\
The pulsation/oscillation is evident both from the phase space portrait in Fig. \ref{fig:phase_ext} and from the fact that in Fig. \ref{fig:sph}, after a full collapsing/expanding cycle, the system comes back at the initial values for all the four variables $(\mathcal{A},\dot{\mathcal{A}},\ddot{\mathcal{A}}, P_\mathcal{A})$ and hence the motion is periodic, with a comoving period of:
 \begin{equation}
  T_{Cyc} = 2\tau_{min} = 4\sqrt{\pi}\Xi p_0\mathcal{I}^{EA}_{+1}(\mathcal{A}_0) 
 \end{equation}
Substituting the value of $\mathcal{I}^{EA}_{+1}(\mathcal{A}_0) $ yields:
\begin{equation}
\label{eqn:puls_time}
T_{Cyc}  = \frac{M_S}{2\Xi p_0}\;\frac{\kappa_p \left(\kappa_S+\kappa_p/2\right)}{\left[\kappa_S \left(\kappa_S+\kappa_p\right)\right]^{3/2}}
\end{equation}
For Sun-like parameters  the comoving time period is of order $10^{-5}\;s$, to which gravitational time dilation should be applied in order to evaluate the cycle period for a distant observer.\\
We will call this new objects periodically radially pulsating objetcs (PRPOs).
\subsection{\label{subsec:quant}Quantum Effects}
\begin{figure}[ht]
\begin{minipage}{0.48\columnwidth}
\centering
	 \includegraphics[scale=1]{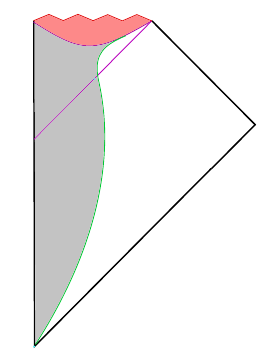}
    \caption{Carter-Penrose Diagram of a Collapsing Black Hole Geometry. In grey the area excluded from the external geometry, in green the sphere surface trajectory, in red the dynamically removed area with $R_S<R_{min}$. The horizontal purple line is the event horizon. }
    \label{fig:penrose_BH}
\end{minipage}
\hfill
\begin{minipage}{0.48\columnwidth}
\centering
	 \includegraphics[scale=1]{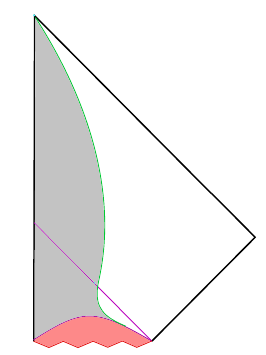}
    \caption{Carter-Penrose Diagram of a Collapsing White Hole Geometry. In grey the area excluded from the external geometry, in green the sphere surface trajectory, in red the dynamically removed area with $R_S<R_{min}$. The horizontal purple line is the event anti-horizon.}
    \label{fig:penrose_WH}
 \end{minipage}
\end{figure}

The dominant effect which arise in the quantum regime, as said in the previous discussion on the inversion behaviour, is that of a quantum transition between the two geometries of a black and a white hole.\\
The two geometries under evaluation are depicted by the Penrose diagrams\cite{Penrose1963} in Fig. \ref{fig:penrose_BH} and Fig. \ref{fig:penrose_WH}. We show only the flat model, since the spherical model cannot be described by a single patch of coordinates at infinity.\\
In the transition we have a sudden change of geometry, but some of the spacetime properties are identified between the two geometries. What we need to do is identify the null-like infinity and the time-like infinity of the two geometries. Also the space-like infinity and the minimal dynamical radius are to be identified. Those identifications are such that the event horizon of the collapsing sphere effectively become an anti-horizon in the out-rising picture, so the out-rising geometry is recognised to be the one of a white hole.\\
This means that, in the quantum regime, the transition actually transforms the horizon, while the asymptotic geometry remains the same.\\
\\
We now try to put some bounds on the transition rate $\Gamma_{B\rightarrow W}$ with some simplified arguments.\\
The probability of seeing a transition from a black to a white hole could be approximated by:
\begin{equation}
    P(t) = 1-e^{-\Gamma_{B\rightarrow W} t}
\end{equation}
Since we know that the rate is the inverse of the mean time for the transition, namely $\Gamma_{B\rightarrow W} = \frac{1}{\langle t \rangle}$, we can infer the transition rate from the knowledge of the mean time.\\
Assuming that we have not seen yet an astrophysical event recognizable as a white hole, we can infer that the mean time for the transition should be greater than the age of the universe, so we obtain:
\begin{equation}
    \langle t \rangle > t_{univ} \Longrightarrow \Gamma_{B\rightarrow W} <  \frac{1}{t_{univ}}
\end{equation}
The Planck 2018 data\cite{Planck2018} given an estimate of $t_{univ} = (13.787 \pm 0.020) 10^9$yrs, so an estimate of the transition rate could be given by:
\begin{equation}
    \Gamma_{B\rightarrow W} \lesssim 7.25 \cdot 10^{-11} \; \text{Events}/\text{yrs}
\end{equation}
This transition rate should be evaluated carefully in a full quantization of the theory, which is beyond the scope of this work. We can only constrain it to the experimental facts that we are confronted with, which state that the probability of such transition needs to be quite low.\\
As also stated in subsubsection \ref{subsubsec:transit}, in a more realistic model is possible that no transition occurs at all if the star is stabilized inside the event horizon, so in that scenario the transition rate $\Gamma_{B\rightarrow W}$ should be zero.
\section{\label{sec:conc}CONCLUSIONS}
In this paper we analyzed the role played by a non-zero area gap in the crucial setting of gravitational collapse.\\
In Section \ref{sec:eff_OS} we have found that, thanks to the polymer tension (\ref{eqn:poly_press}) caused by the discrete nature of the area, the collapse is regularized both in the flat and in the spherical case, avoiding the formation of a singularity. This implies the existence of a dynamical minimum surface that the sphere can reach, given by equations (\ref{eqn:maxmin_p1_poly} -  \ref{eqn:maxmin_p0_poly}). In equations (\ref{eqn:min_outside}) we showed that exist values of the polymer parameter for which said minima lie outside of the event horizon. Those are compatible with the subluminal collapsing speed constraint imposed on $\mu_0$ in equation (\ref{eqn:poly_subl}).\\
In Section \ref{sec:phen} we characterized such polymer tension, and we have found that a pressure of quantum origin has macroscopic consequences for the collapse, so conceptually it is not very different from the various degeneracy pressures which appear in the usual treatment of compact objects.\\
We have also found that on the new regular minima there is a velocity inversion, which has two important consequences depending on whether or not it happens inside or outside the event horizon.\\
We found that when such an inversion happens inside the event horizon, the associated process has to be  understood as a quantum transition from a black hole geometry to a white hole geometry, in accordance with the literature cited. We also considered the possibility that such process is avoided completely in a more realistic star model, and this is more plausible according to our astrophysical data.\\
More work is needed in order to implement the polymer/LQG procedure for models with rotating stars or internal pressure, which could provide a mechanism to dissipate the quantum gravitational pressure into internal degrees of freedom of the star and stabilize it without the formation of a white hole.\\
We then found the conditions for the bounce to happen outside the event horizon, finding a new class of objects, PRPOs, which exhibit a periodic radially pulsating behaviour, with the comoving pulsation period given by equation (\ref{eqn:puls_time}). \\
We state that those objects, due to the presence of the new polymer tension, could have a different deformability with respect to neutron stars or black holes, and hence they could leave a different signature in gravitational waves emitted by binary mergers.\\
A careful study of such signatures, in the case of more accurate star models, is left for future investigations in the hope for new tests of quantum gravity.
\begin{acknowledgments}
We wolud like to thank Paolo Pani for fruitful discussions on this subject.
\end{acknowledgments}

\appendix

\section{\label{app:integr}Integrals of the Equations of Motion}
We make use of this Appendix to define the integrals defined throughout the sections \ref{sec:ham_OS} and \ref{sec:eff_OS}.\\
\subsection{Classical Dynamics Integrals}
\subsubsection{Surface Integrals}
We start with the classical integrals for the surface equations of motion.\\
Let's define the classical area (CA) integrals as:
\begin{equation}
\label{eqn:int_eom_cla_S}
\mathcal{I}_\epsilon^{CA}(\mathcal{A})= \int{\frac{d\mathcal{A}}{ \sqrt{4M_S  \sqrt{\pi\mathcal{A}}-\kappa_S\mathcal{A}}}}
\end{equation}
Those are two integrals, one for each value of $\epsilon$, and are equivalent to:
\begin{align}
\label{eqn:int_eom_cla_S_0}
\mathcal{I}_{0}^{CA}(\mathcal{A}) =& \frac{2\mathcal{A}}{3  \sqrt{M_S  \sqrt{\pi\mathcal{A}}}}\\
\begin{split}
\label{eqn:int_eom_cla_S_+1}
\mathcal{I}_{+1}^{CA}(\mathcal{A}) =& \frac{8M_S\sqrt{\pi } \; \text{atan}\left(\frac{\sqrt{\kappa_S\mathcal{A}}}{\sqrt{4 M_S \sqrt{\pi\mathcal{A}}-\kappa_S\mathcal{A}}}\right)}{\left(\kappa_S\right)^{3/2}}\\
&-\frac{2\sqrt{4M_S\sqrt{\pi\mathcal{A}}-\kappa_S\mathcal{A}}}{\kappa_S}
\end{split}
\end{align}
Those can be directly put into the solutions of the equations of motion.
\subsubsection{Momentum Integrals}
Let's define the classical momentum (CM) integrals as:
\begin{equation}
\mathcal{I}_\epsilon^{CM}(P_{\mathcal{A}})= \int\frac{256 P_{\mathcal{A}} d P_{\mathcal{A}}}{\frac{\kappa_S}{\pi ^2 \Xi ^2}+64 P_{\mathcal{A}}^2}
\end{equation}
Those are two integrals, one for each value of $\epsilon$, evaluated to be:
\begin{align}
\mathcal{I}_{0}^{CM}(P_{\mathcal{A}}) &= 4 \ln\left(P_{\mathcal{A}}\right)\\
\mathcal{I}_{+1}^{CM}(P_{\mathcal{A}}) &= 2 \ln\left(\kappa_S+64 \pi ^2 \Xi ^2  P_{\mathcal{A}}^2\right)
\end{align}
The integral in the area variable instead yields:
\begin{equation}
    -\int \frac{d\mathcal{A}}{\mathcal{A}} = -\ln\left(\mathcal{A}\right)
\end{equation}
Plugging those integrals in the classical momentum equations we can invert the logarithms to obtain $P_\mathcal{A}(\mathcal{A})$.\\
\begin{widetext}
\subsection{Effective Dynamics Integrals}
\subsubsection{Surface Integrals}
We continue this appendix with the effective integrals for the surface equations of motion.\\
Let's define the effective area (EA) integrals as:
\begin{equation}
\begin{split}
&\mathcal{I}^{EA}_\epsilon(\mathcal{A}) =\int{\frac{ d\mathcal{A}}{\sqrt{\left(4M_S\sqrt{\pi } -\kappa_S \sqrt{\mathcal{A}}\right) \left(\sqrt{\mathcal{A}} \left(\kappa_S +\kappa_p\right)-4 M_S\sqrt{\pi}\right)}}}
\end{split}
\end{equation}
That evaluates to:
\begin{equation}
\mathcal{I}_0^{EA}(\mathcal{A}) = \frac{ \left(M_S+8 \pi ^{3/2} p_0^2 \sqrt{\mathcal{A}}\right)}{384 \pi ^{7/2}  p_0^4}\sqrt{\frac{\left(16 \pi ^{3/2} p_0^2 \sqrt{\mathcal{A}}-M_S \right)}{M_S}}
\end{equation}
\begin{equation}
\begin{split}
\mathcal{I}_{+1}^{EA}(\mathcal{A}) &=\frac{16 M_S \sqrt{\pi } \left(\kappa_S+\kappa_p/2\right) \text{arctan}\left(\frac{\sqrt{\kappa_S \left(\sqrt{\mathcal{A}} \left(\kappa_S+\kappa_p\right)-4 M_S \sqrt{\pi }  \right)}}{\sqrt{-\left(\kappa_S+\kappa_p\right) \left(\kappa_S \sqrt{\mathcal{A}}-4 M_S\sqrt{\pi } \right)}}\right)}{\left[\kappa_S \left(\kappa_S+\kappa_p\right)\right]^{3/2}}-2\frac{\sqrt{\left(4M_S\sqrt{\pi } -\kappa_S\sqrt{\mathcal{A}}\right) \left(\sqrt{\mathcal{A}} \left(\kappa_S+\kappa_p\right)-4M_S\sqrt{\pi } \right)}}{\kappa_S \left(\kappa_S+\kappa_p\right)}
\end{split}
\end{equation}
We got two different integrals, just like before, since those evaluate to different functions for different $\epsilon$ parameters.
\end{widetext}

\bibliography{biblio}

\begin{thebibliography}{48}%
\makeatletter
\providecommand \@ifxundefined [1]{%
 \@ifx{#1\undefined}
}%
\providecommand \@ifnum [1]{%
 \ifnum #1\expandafter \@firstoftwo
 \else \expandafter \@secondoftwo
 \fi
}%
\providecommand \@ifx [1]{%
 \ifx #1\expandafter \@firstoftwo
 \else \expandafter \@secondoftwo
 \fi
}%
\providecommand \natexlab [1]{#1}%
\providecommand \enquote  [1]{``#1''}%
\providecommand \bibnamefont  [1]{#1}%
\providecommand \bibfnamefont [1]{#1}%
\providecommand \citenamefont [1]{#1}%
\providecommand \href@noop [0]{\@secondoftwo}%
\providecommand \href [0]{\begingroup \@sanitize@url \@href}%
\providecommand \@href[1]{\@@startlink{#1}\@@href}%
\providecommand \@@href[1]{\endgroup#1\@@endlink}%
\providecommand \@sanitize@url [0]{\catcode `\\12\catcode `\$12\catcode
  `\&12\catcode `\#12\catcode `\^12\catcode `\_12\catcode `\%12\relax}%
\providecommand \@@startlink[1]{}%
\providecommand \@@endlink[0]{}%
\providecommand \url  [0]{\begingroup\@sanitize@url \@url }%
\providecommand \@url [1]{\endgroup\@href {#1}{\urlprefix }}%
\providecommand \urlprefix  [0]{URL }%
\providecommand \Eprint [0]{\href }%
\providecommand \doibase [0]{https://doi.org/}%
\providecommand \selectlanguage [0]{\@gobble}%
\providecommand \bibinfo  [0]{\@secondoftwo}%
\providecommand \bibfield  [0]{\@secondoftwo}%
\providecommand \translation [1]{[#1]}%
\providecommand \BibitemOpen [0]{}%
\providecommand \bibitemStop [0]{}%
\providecommand \bibitemNoStop [0]{.\EOS\space}%
\providecommand \EOS [0]{\spacefactor3000\relax}%
\providecommand \BibitemShut  [1]{\csname bibitem#1\endcsname}%
\let\auto@bib@innerbib\@empty
\bibitem [{\citenamefont {Misner}\ \emph {et~al.}(1973)\citenamefont {Misner},
  \citenamefont {Thorne},\ and\ \citenamefont {Wheeler}}]{Misner:1973prb}%
  \BibitemOpen
  \bibfield  {author} {\bibinfo {author} {\bibfnamefont {C.~W.}\ \bibnamefont
  {Misner}}, \bibinfo {author} {\bibfnamefont {K.~S.}\ \bibnamefont {Thorne}},\
  and\ \bibinfo {author} {\bibfnamefont {J.~A.}\ \bibnamefont {Wheeler}},\
  }\href@noop {} {\emph {\bibinfo {title} {{Gravitation}}}}\ (\bibinfo
  {publisher} {W. H. Freeman},\ \bibinfo {address} {San Francisco},\ \bibinfo
  {year} {1973})\BibitemShut {NoStop}%
\bibitem [{\citenamefont {Weinberg}(2008)}]{Weinberg2008}%
  \BibitemOpen
  \bibfield  {author} {\bibinfo {author} {\bibfnamefont {S.}~\bibnamefont
  {Weinberg}},\ }\href@noop {} {\emph {\bibinfo {title} {Cosmology}}}\
  (\bibinfo  {publisher} {Oxford University Press},\ \bibinfo {address}
  {London, England},\ \bibinfo {year} {2008})\BibitemShut {NoStop}%
\bibitem [{\citenamefont {Shapiro}\ and\ \citenamefont
  {Teukolsky}(1983)}]{Shapiro1983-fp}%
  \BibitemOpen
  \bibfield  {author} {\bibinfo {author} {\bibfnamefont {S.~L.}\ \bibnamefont
  {Shapiro}}\ and\ \bibinfo {author} {\bibfnamefont {S.~A.}\ \bibnamefont
  {Teukolsky}},\ }\href@noop {} {\emph {\bibinfo {title} {Black holes, white
  dwarfs, and neutron stars}}}\ (\bibinfo  {publisher} {John Wiley \& Sons},\
  \bibinfo {address} {Nashville, TN},\ \bibinfo {year} {1983})\BibitemShut
  {NoStop}%
\bibitem [{\citenamefont {Kalogera}\ and\ \citenamefont
  {Baym}(1996)}]{Kalogera_1996}%
  \BibitemOpen
  \bibfield  {author} {\bibinfo {author} {\bibfnamefont {V.}~\bibnamefont
  {Kalogera}}\ and\ \bibinfo {author} {\bibfnamefont {G.}~\bibnamefont
  {Baym}},\ }\bibfield  {title} {\bibinfo {title} {The maximum mass of a
  neutron star},\ }\href {https://doi.org/10.1086/310296} {\bibfield  {journal}
  {\bibinfo  {journal} {The Astrophysical Journal}\ }\textbf {\bibinfo {volume}
  {470}},\ \bibinfo {pages} {L61} (\bibinfo {year} {1996})}\BibitemShut
  {NoStop}%
\bibitem [{\citenamefont {Rezzolla}\ \emph {et~al.}(2018)\citenamefont
  {Rezzolla}, \citenamefont {Most},\ and\ \citenamefont
  {Weih}}]{Rezzolla_2018}%
  \BibitemOpen
  \bibfield  {author} {\bibinfo {author} {\bibfnamefont {L.}~\bibnamefont
  {Rezzolla}}, \bibinfo {author} {\bibfnamefont {E.~R.}\ \bibnamefont {Most}},\
  and\ \bibinfo {author} {\bibfnamefont {L.~R.}\ \bibnamefont {Weih}},\
  }\bibfield  {title} {\bibinfo {title} {Using gravitational-wave observations
  and quasi-universal relations to constrain the maximum mass of neutron
  stars},\ }\href {https://doi.org/10.3847/2041-8213/aaa401} {\bibfield
  {journal} {\bibinfo  {journal} {The Astrophysical Journal Letters}\ }\textbf
  {\bibinfo {volume} {852}},\ \bibinfo {pages} {L25} (\bibinfo {year}
  {2018})}\BibitemShut {NoStop}%
\bibitem [{\citenamefont {DeWitt}(1967)}]{WdW67}%
  \BibitemOpen
  \bibfield  {author} {\bibinfo {author} {\bibfnamefont {B.~S.}\ \bibnamefont
  {DeWitt}},\ }\bibfield  {title} {\bibinfo {title} {Quantum theory of gravity.
  i. the canonical theory},\ }\href {https://doi.org/10.1103/PhysRev.160.1113}
  {\bibfield  {journal} {\bibinfo  {journal} {Phys. Rev.}\ }\textbf {\bibinfo
  {volume} {160}},\ \bibinfo {pages} {1113} (\bibinfo {year}
  {1967})}\BibitemShut {NoStop}%
\bibitem [{\citenamefont {Thiemann}(2007)}]{Thiemann_2007}%
  \BibitemOpen
  \bibfield  {author} {\bibinfo {author} {\bibfnamefont {T.}~\bibnamefont
  {Thiemann}},\ }\href@noop {} {\emph {\bibinfo {title} {Modern Canonical
  Quantum General Relativity}}}\ (\bibinfo  {publisher} {Cambridge University
  Press},\ \bibinfo {year} {2007})\BibitemShut {NoStop}%
\bibitem [{\citenamefont {Oppenheimer}\ and\ \citenamefont
  {Snyder}(1939)}]{Oppenheimer_Snyder_1939}%
  \BibitemOpen
  \bibfield  {author} {\bibinfo {author} {\bibfnamefont {J.~R.}\ \bibnamefont
  {Oppenheimer}}\ and\ \bibinfo {author} {\bibfnamefont {H.}~\bibnamefont
  {Snyder}},\ }\bibfield  {title} {\bibinfo {title} {On continued gravitational
  contraction},\ }\href {https://doi.org/10.1103/PhysRev.56.455} {\bibfield
  {journal} {\bibinfo  {journal} {Phys. Rev.}\ }\textbf {\bibinfo {volume}
  {56}},\ \bibinfo {pages} {455} (\bibinfo {year} {1939})}\BibitemShut
  {NoStop}%
\bibitem [{\citenamefont {Corichi}\ \emph {et~al.}(2007)\citenamefont
  {Corichi}, \citenamefont {Vuka\ifmmode~\check{s}\else \v{s}\fi{}inac},\ and\
  \citenamefont {Zapata}}]{Corichi_2007}%
  \BibitemOpen
  \bibfield  {author} {\bibinfo {author} {\bibfnamefont {A.}~\bibnamefont
  {Corichi}}, \bibinfo {author} {\bibfnamefont {T.}~\bibnamefont
  {Vuka\ifmmode~\check{s}\else \v{s}\fi{}inac}},\ and\ \bibinfo {author}
  {\bibfnamefont {J.~A.}\ \bibnamefont {Zapata}},\ }\bibfield  {title}
  {\bibinfo {title} {Polymer quantum mechanics and its continuum limit},\
  }\href {https://doi.org/10.1103/PhysRevD.76.044016} {\bibfield  {journal}
  {\bibinfo  {journal} {Phys. Rev. D}\ }\textbf {\bibinfo {volume} {76}},\
  \bibinfo {pages} {044016} (\bibinfo {year} {2007})}\BibitemShut {NoStop}%
\bibitem [{\citenamefont {Hájíček}\ and\ \citenamefont
  {Kiefer}(2001)}]{sing_kief}%
  \BibitemOpen
  \bibfield  {author} {\bibinfo {author} {\bibfnamefont {P.}~\bibnamefont
  {Hájíček}}\ and\ \bibinfo {author} {\bibfnamefont {C.}~\bibnamefont
  {Kiefer}},\ }\bibfield  {title} {\bibinfo {title} {Singularity avoidance by
  collapsing shells in quantum gravity},\ }\href
  {https://doi.org/10.1142/S0218271801001578} {\bibfield  {journal} {\bibinfo
  {journal} {International Journal of Modern Physics D}\ }\textbf {\bibinfo
  {volume} {10}},\ \bibinfo {pages} {775} (\bibinfo {year} {2001})}\BibitemShut
  {NoStop}%
\bibitem [{\citenamefont {Kiefer}\ and\ \citenamefont
  {Schmitz}(2019)}]{LTB_Kief}%
  \BibitemOpen
  \bibfield  {author} {\bibinfo {author} {\bibfnamefont {C.}~\bibnamefont
  {Kiefer}}\ and\ \bibinfo {author} {\bibfnamefont {T.}~\bibnamefont
  {Schmitz}},\ }\bibfield  {title} {\bibinfo {title} {Singularity avoidance for
  collapsing quantum dust in the lema\^{\i}tre-tolman-bondi model},\ }\href
  {https://doi.org/10.1103/PhysRevD.99.126010} {\bibfield  {journal} {\bibinfo
  {journal} {Phys. Rev. D}\ }\textbf {\bibinfo {volume} {99}},\ \bibinfo
  {pages} {126010} (\bibinfo {year} {2019})}\BibitemShut {NoStop}%
\bibitem [{\citenamefont {Malafarina}(2017)}]{Malafrina}%
  \BibitemOpen
  \bibfield  {author} {\bibinfo {author} {\bibfnamefont {D.}~\bibnamefont
  {Malafarina}},\ }\bibfield  {title} {\bibinfo {title} {Classical collapse to
  black holes and quantum bounces: A review},\ }\bibfield  {journal} {\bibinfo
  {journal} {Universe}\ }\textbf {\bibinfo {volume} {3}},\ \href
  {https://doi.org/10.3390/universe3020048} {10.3390/universe3020048} (\bibinfo
  {year} {2017})\BibitemShut {NoStop}%
\bibitem [{\citenamefont {Gourgoulhon}(2012)}]{Gourgoulhon:2007ue}%
  \BibitemOpen
  \bibfield  {author} {\bibinfo {author} {\bibfnamefont {E.}~\bibnamefont
  {Gourgoulhon}},\ }\href {https://doi.org/10.1007/978-3-642-24525-1} {\emph
  {\bibinfo {title} {3+1 Formalism in General Relativity: Bases of Numerical
  Relativity}}}\ (\bibinfo  {publisher} {Springer Berlin Heidelberg},\ \bibinfo
  {year} {2012})\BibitemShut {NoStop}%
\bibitem [{\citenamefont {Arnowitt}\ \emph {et~al.}(1959)\citenamefont
  {Arnowitt}, \citenamefont {Deser},\ and\ \citenamefont {Misner}}]{ADM_1959}%
  \BibitemOpen
  \bibfield  {author} {\bibinfo {author} {\bibfnamefont {R.}~\bibnamefont
  {Arnowitt}}, \bibinfo {author} {\bibfnamefont {S.}~\bibnamefont {Deser}},\
  and\ \bibinfo {author} {\bibfnamefont {C.~W.}\ \bibnamefont {Misner}},\
  }\bibfield  {title} {\bibinfo {title} {Dynamical structure and definition of
  energy in general relativity},\ }\href
  {https://doi.org/10.1103/PhysRev.116.1322} {\bibfield  {journal} {\bibinfo
  {journal} {Phys. Rev.}\ }\textbf {\bibinfo {volume} {116}},\ \bibinfo {pages}
  {1322} (\bibinfo {year} {1959})}\BibitemShut {NoStop}%
\bibitem [{\citenamefont {Poisson}(2004)}]{Poisson_2004}%
  \BibitemOpen
  \bibfield  {author} {\bibinfo {author} {\bibfnamefont {E.}~\bibnamefont
  {Poisson}},\ }\href@noop {} {\emph {\bibinfo {title} {A Relativist’s
  Toolkit: The Mathematics of Black-Hole Mechanics}}}\ (\bibinfo  {publisher}
  {Cambridge University Press},\ \bibinfo {year} {2004})\BibitemShut {NoStop}%
\bibitem [{\citenamefont {H\'aj\'{\i}\ifmmode~\check{c}\else \v{c}\fi{}ek}\
  and\ \citenamefont {Kijowski}(1998)}]{Hájíček_Kijowski_1998}%
  \BibitemOpen
  \bibfield  {author} {\bibinfo {author} {\bibfnamefont {P.}~\bibnamefont
  {H\'aj\'{\i}\ifmmode~\check{c}\else \v{c}\fi{}ek}}\ and\ \bibinfo {author}
  {\bibfnamefont {J.}~\bibnamefont {Kijowski}},\ }\bibfield  {title} {\bibinfo
  {title} {Lagrangian and hamiltonian formalism for discontinuous fluid and
  gravitational field},\ }\href {https://doi.org/10.1103/PhysRevD.57.914}
  {\bibfield  {journal} {\bibinfo  {journal} {Phys. Rev. D}\ }\textbf {\bibinfo
  {volume} {57}},\ \bibinfo {pages} {914} (\bibinfo {year} {1998})}\BibitemShut
  {NoStop}%
\bibitem [{\citenamefont {Kanai}\ \emph {et~al.}(2011)\citenamefont {Kanai},
  \citenamefont {Siino},\ and\ \citenamefont {Hosoya}}]{PG_coord_Kanai}%
  \BibitemOpen
  \bibfield  {author} {\bibinfo {author} {\bibfnamefont {Y.}~\bibnamefont
  {Kanai}}, \bibinfo {author} {\bibfnamefont {M.}~\bibnamefont {Siino}},\ and\
  \bibinfo {author} {\bibfnamefont {A.}~\bibnamefont {Hosoya}},\ }\bibfield
  {title} {\bibinfo {title} {{Gravitational Collapse in Painlevé-Gullstrand
  Coordinates}},\ }\href {https://doi.org/10.1143/PTP.125.1053} {\bibfield
  {journal} {\bibinfo  {journal} {Progress of Theoretical Physics}\ }\textbf
  {\bibinfo {volume} {125}},\ \bibinfo {pages} {1053} (\bibinfo {year}
  {2011})}\BibitemShut {NoStop}%
\bibitem [{\citenamefont {Casadio}(1998)}]{Casadio_1998}%
  \BibitemOpen
  \bibfield  {author} {\bibinfo {author} {\bibfnamefont {R.}~\bibnamefont
  {Casadio}},\ }\bibfield  {title} {\bibinfo {title} {Hamiltonian formalism for
  the oppenheimer-snyder model},\ }\href
  {https://doi.org/10.1103/PhysRevD.58.064013} {\bibfield  {journal} {\bibinfo
  {journal} {Phys. Rev. D}\ }\textbf {\bibinfo {volume} {58}},\ \bibinfo
  {pages} {064013} (\bibinfo {year} {1998})}\BibitemShut {NoStop}%
\bibitem [{\citenamefont {Kiefer}\ and\ \citenamefont
  {Mohaddes}(2023)}]{Kiefer_Mohaddes_2023}%
  \BibitemOpen
  \bibfield  {author} {\bibinfo {author} {\bibfnamefont {C.}~\bibnamefont
  {Kiefer}}\ and\ \bibinfo {author} {\bibfnamefont {H.}~\bibnamefont
  {Mohaddes}},\ }\bibfield  {title} {\bibinfo {title} {From classical to
  quantum oppenheimer-snyder model: Nonmarginal case},\ }\href
  {https://doi.org/10.1103/PhysRevD.107.126006} {\bibfield  {journal} {\bibinfo
   {journal} {Phys. Rev. D}\ }\textbf {\bibinfo {volume} {107}},\ \bibinfo
  {pages} {126006} (\bibinfo {year} {2023})}\BibitemShut {NoStop}%
\bibitem [{\citenamefont {Schmitz}(2020)}]{Schmitz_2020}%
  \BibitemOpen
  \bibfield  {author} {\bibinfo {author} {\bibfnamefont {T.}~\bibnamefont
  {Schmitz}},\ }\bibfield  {title} {\bibinfo {title} {Towards a quantum
  oppenheimer-snyder model},\ }\href
  {https://doi.org/10.1103/PhysRevD.101.026016} {\bibfield  {journal} {\bibinfo
   {journal} {Phys. Rev. D}\ }\textbf {\bibinfo {volume} {101}},\ \bibinfo
  {pages} {026016} (\bibinfo {year} {2020})}\BibitemShut {NoStop}%
\bibitem [{\citenamefont {Brown}\ and\ \citenamefont
  {Kucha\ifmmode~\check{r}\else \v{r}\fi{}}(1995)}]{Brown_Kuchař_1995}%
  \BibitemOpen
  \bibfield  {author} {\bibinfo {author} {\bibfnamefont {J.~D.}\ \bibnamefont
  {Brown}}\ and\ \bibinfo {author} {\bibfnamefont {K.~V.}\ \bibnamefont
  {Kucha\ifmmode~\check{r}\else \v{r}\fi{}}},\ }\bibfield  {title} {\bibinfo
  {title} {Dust as a standard of space and time in canonical quantum gravity},\
  }\href {https://doi.org/10.1103/PhysRevD.51.5600} {\bibfield  {journal}
  {\bibinfo  {journal} {Phys. Rev. D}\ }\textbf {\bibinfo {volume} {51}},\
  \bibinfo {pages} {5600} (\bibinfo {year} {1995})}\BibitemShut {NoStop}%
\bibitem [{\citenamefont {Painlevé}(1921)}]{Painleve_1921}%
  \BibitemOpen
  \bibfield  {author} {\bibinfo {author} {\bibfnamefont {P.}~\bibnamefont
  {Painlevé}},\ }\bibfield  {title} {\bibinfo {title} {La mécanique classique
  et la théorie de la relativité},\ }\href@noop {} {\bibfield  {journal}
  {\bibinfo  {journal} {C. R. Acad. Sci.}\ }\textbf {\bibinfo {volume} {173}},\
  \bibinfo {pages} {677} (\bibinfo {year} {1921})}\BibitemShut {NoStop}%
\bibitem [{\citenamefont {Gullstrand}(1922)}]{Gullstrand:1922tfa}%
  \BibitemOpen
  \bibfield  {author} {\bibinfo {author} {\bibfnamefont {A.}~\bibnamefont
  {Gullstrand}},\ }\href@noop {} {\emph {\bibinfo {title} {{Allgemeine L\"osung
  des statischen Eink\"orperproblems in der Einsteinschen
  Gravitationstheorie}}}},\ \bibinfo {series} {Arkiv f\"or matematik, astronomi
  och fysik}, Vol.\ \bibinfo {volume} {16,8}\ (\bibinfo  {publisher} {Almqvist
  \& Wiksell},\ \bibinfo {address} {Stockholm},\ \bibinfo {year}
  {1922})\BibitemShut {NoStop}%
\bibitem [{\citenamefont {Gautreau}\ and\ \citenamefont
  {Hoffmann}(1978)}]{Gautreau_Hoffmann_1978}%
  \BibitemOpen
  \bibfield  {author} {\bibinfo {author} {\bibfnamefont {R.}~\bibnamefont
  {Gautreau}}\ and\ \bibinfo {author} {\bibfnamefont {B.}~\bibnamefont
  {Hoffmann}},\ }\bibfield  {title} {\bibinfo {title} {The schwarzschild radial
  coordinate as a measure of proper distance},\ }\href
  {https://doi.org/10.1103/PhysRevD.17.2552} {\bibfield  {journal} {\bibinfo
  {journal} {Phys. Rev. D}\ }\textbf {\bibinfo {volume} {17}},\ \bibinfo
  {pages} {2552} (\bibinfo {year} {1978})}\BibitemShut {NoStop}%
\bibitem [{\citenamefont {Strocchi}(2016)}]{Strocchi_2016}%
  \BibitemOpen
  \bibfield  {author} {\bibinfo {author} {\bibfnamefont {F.}~\bibnamefont
  {Strocchi}},\ }\href {https://doi.org/10.1007/978-3-319-17695-6} {\emph
  {\bibinfo {title} {Gauge Invariance and Weyl-polymer Quantization}}}\
  (\bibinfo  {publisher} {Springer International Publishing},\ \bibinfo {year}
  {2016})\BibitemShut {NoStop}%
\bibitem [{\citenamefont {Morchio}\ and\ \citenamefont
  {Strocchi}(2007)}]{Morchio_2007}%
  \BibitemOpen
  \bibfield  {author} {\bibinfo {author} {\bibfnamefont {G.}~\bibnamefont
  {Morchio}}\ and\ \bibinfo {author} {\bibfnamefont {F.}~\bibnamefont
  {Strocchi}},\ }\bibfield  {title} {\bibinfo {title} {{Quantum mechanics on
  manifolds and topological effects}},\ }\href
  {https://doi.org/10.1007/s11005-007-0188-5} {\bibfield  {journal} {\bibinfo
  {journal} {Lett. Math. Phys.}\ }\textbf {\bibinfo {volume} {82}},\ \bibinfo
  {pages} {219} (\bibinfo {year} {2007})}\BibitemShut {NoStop}%
\bibitem [{\citenamefont {Rovelli}\ and\ \citenamefont
  {Smolin}(1995{\natexlab{a}})}]{Rovelli_1994}%
  \BibitemOpen
  \bibfield  {author} {\bibinfo {author} {\bibfnamefont {C.}~\bibnamefont
  {Rovelli}}\ and\ \bibinfo {author} {\bibfnamefont {L.}~\bibnamefont
  {Smolin}},\ }\bibfield  {title} {\bibinfo {title} {{Discreteness of area and
  volume in quantum gravity}},\ }\href
  {https://doi.org/10.1016/0550-3213(95)00150-Q} {\bibfield  {journal}
  {\bibinfo  {journal} {Nucl. Phys. B}\ }\textbf {\bibinfo {volume} {442}},\
  \bibinfo {pages} {593} (\bibinfo {year} {1995}{\natexlab{a}})},\ \bibinfo
  {note} {[Erratum: Nucl.Phys.B 456, 753--754 (1995)]}\BibitemShut {NoStop}%
\bibitem [{\citenamefont {Rovelli}\ and\ \citenamefont
  {Smolin}(1995{\natexlab{b}})}]{Rovelli_1995}%
  \BibitemOpen
  \bibfield  {author} {\bibinfo {author} {\bibfnamefont {C.}~\bibnamefont
  {Rovelli}}\ and\ \bibinfo {author} {\bibfnamefont {L.}~\bibnamefont
  {Smolin}},\ }\bibfield  {title} {\bibinfo {title} {{Spin networks and quantum
  gravity}},\ }\href {https://doi.org/10.1103/PhysRevD.52.5743} {\bibfield
  {journal} {\bibinfo  {journal} {Phys. Rev. D}\ }\textbf {\bibinfo {volume}
  {52}},\ \bibinfo {pages} {5743} (\bibinfo {year}
  {1995}{\natexlab{b}})}\BibitemShut {NoStop}%
\bibitem [{\citenamefont {Ashtekar}\ \emph {et~al.}(2003)\citenamefont
  {Ashtekar}, \citenamefont {Fairhurst},\ and\ \citenamefont
  {Willis}}]{Ashtekar_2002}%
  \BibitemOpen
  \bibfield  {author} {\bibinfo {author} {\bibfnamefont {A.}~\bibnamefont
  {Ashtekar}}, \bibinfo {author} {\bibfnamefont {S.}~\bibnamefont
  {Fairhurst}},\ and\ \bibinfo {author} {\bibfnamefont {J.~L.}\ \bibnamefont
  {Willis}},\ }\bibfield  {title} {\bibinfo {title} {{Quantum gravity, shadow
  states, and quantum mechanics}},\ }\href
  {https://doi.org/10.1088/0264-9381/20/6/302} {\bibfield  {journal} {\bibinfo
  {journal} {Class. Quant. Grav.}\ }\textbf {\bibinfo {volume} {20}},\ \bibinfo
  {pages} {1031} (\bibinfo {year} {2003})}\BibitemShut {NoStop}%
\bibitem [{\citenamefont {Haggard}\ and\ \citenamefont
  {Rovelli}(2015)}]{Haggard2015}%
  \BibitemOpen
  \bibfield  {author} {\bibinfo {author} {\bibfnamefont {H.~M.}\ \bibnamefont
  {Haggard}}\ and\ \bibinfo {author} {\bibfnamefont {C.}~\bibnamefont
  {Rovelli}},\ }\bibfield  {title} {\bibinfo {title} {Quantum-gravity effects
  outside the horizon spark black to white hole tunneling},\ }\bibfield
  {journal} {\bibinfo  {journal} {Physical Review D}\ }\textbf {\bibinfo
  {volume} {92}},\ \href {https://doi.org/10.1103/physrevd.92.104020}
  {10.1103/physrevd.92.104020} (\bibinfo {year} {2015})\BibitemShut {NoStop}%
\bibitem [{\citenamefont {M\"{u}nch}(2021)}]{Mnch2021}%
  \BibitemOpen
  \bibfield  {author} {\bibinfo {author} {\bibfnamefont {J.}~\bibnamefont
  {M\"{u}nch}},\ }\bibfield  {title} {\bibinfo {title} {Effective quantum dust
  collapse via surface matching},\ }\bibfield  {journal} {\bibinfo  {journal}
  {Classical and Quantum Gravity}\ }\textbf {\bibinfo {volume} {38}},\ \href
  {https://doi.org/10.1088/1361-6382/ac103e} {10.1088/1361-6382/ac103e}
  (\bibinfo {year} {2021})\BibitemShut {NoStop}%
\bibitem [{\citenamefont {Kelly}\ \emph {et~al.}(2020)\citenamefont {Kelly},
  \citenamefont {Santacruz},\ and\ \citenamefont {Wilson-Ewing}}]{Kelly2020}%
  \BibitemOpen
  \bibfield  {author} {\bibinfo {author} {\bibfnamefont {J.~G.}\ \bibnamefont
  {Kelly}}, \bibinfo {author} {\bibfnamefont {R.}~\bibnamefont {Santacruz}},\
  and\ \bibinfo {author} {\bibfnamefont {E.}~\bibnamefont {Wilson-Ewing}},\
  }\bibfield  {title} {\bibinfo {title} {Black hole collapse and bounce in
  effective loop quantum gravity},\ }\bibfield  {journal} {\bibinfo  {journal}
  {Classical and Quantum Gravity}\ }\textbf {\bibinfo {volume} {38}},\ \href
  {https://doi.org/10.1088/1361-6382/abd3e2} {10.1088/1361-6382/abd3e2}
  (\bibinfo {year} {2020})\BibitemShut {NoStop}%
\bibitem [{\citenamefont {Bianchi}\ \emph {et~al.}(2018)\citenamefont
  {Bianchi}, \citenamefont {Christodoulou}, \citenamefont {D’Ambrosio},
  \citenamefont {Haggard},\ and\ \citenamefont {Rovelli}}]{Bianchi2018}%
  \BibitemOpen
  \bibfield  {author} {\bibinfo {author} {\bibfnamefont {E.}~\bibnamefont
  {Bianchi}}, \bibinfo {author} {\bibfnamefont {M.}~\bibnamefont
  {Christodoulou}}, \bibinfo {author} {\bibfnamefont {F.}~\bibnamefont
  {D’Ambrosio}}, \bibinfo {author} {\bibfnamefont {H.~M.}\ \bibnamefont
  {Haggard}},\ and\ \bibinfo {author} {\bibfnamefont {C.}~\bibnamefont
  {Rovelli}},\ }\bibfield  {title} {\bibinfo {title} {White holes as remnants:
  a surprising scenario for the end of a black hole},\ }\bibfield  {journal}
  {\bibinfo  {journal} {Classical and Quantum Gravity}\ }\textbf {\bibinfo
  {volume} {35}},\ \href {https://doi.org/10.1088/1361-6382/aae550}
  {10.1088/1361-6382/aae550} (\bibinfo {year} {2018})\BibitemShut {NoStop}%
\bibitem [{\citenamefont {Achour}\ \emph
  {et~al.}(2020{\natexlab{a}})\citenamefont {Achour}, \citenamefont {Brahma},
  \citenamefont {Mukohyama},\ and\ \citenamefont {Uzan}}]{Achour2020}%
  \BibitemOpen
  \bibfield  {author} {\bibinfo {author} {\bibfnamefont {J.~B.}\ \bibnamefont
  {Achour}}, \bibinfo {author} {\bibfnamefont {S.}~\bibnamefont {Brahma}},
  \bibinfo {author} {\bibfnamefont {S.}~\bibnamefont {Mukohyama}},\ and\
  \bibinfo {author} {\bibfnamefont {J.-P.}\ \bibnamefont {Uzan}},\ }\bibfield
  {title} {\bibinfo {title} {Towards consistent black-to-white hole bounces
  from matter collapse},\ }\href
  {https://doi.org/10.1088/1475-7516/2020/09/020} {\bibfield  {journal}
  {\bibinfo  {journal} {Journal of Cosmology and Astroparticle Physics}\
  }\textbf {\bibinfo {volume} {2020}}\bibinfo  {number} { (09)}}\BibitemShut
  {NoStop}%
\bibitem [{\citenamefont {Rovelli}\ and\ \citenamefont
  {Vidotto}(2018)}]{Rovelli_2018}%
  \BibitemOpen
\bibfield  {number} {  }\bibfield  {author} {\bibinfo {author} {\bibfnamefont
  {C.}~\bibnamefont {Rovelli}}\ and\ \bibinfo {author} {\bibfnamefont
  {F.}~\bibnamefont {Vidotto}},\ }\bibfield  {title} {\bibinfo {title} {Small
  black/white hole stability and dark matter},\ }\bibfield  {journal} {\bibinfo
   {journal} {Universe}\ }\textbf {\bibinfo {volume} {4}},\ \href
  {https://doi.org/10.3390/universe4110127} {10.3390/universe4110127} (\bibinfo
  {year} {2018})\BibitemShut {NoStop}%
\bibitem [{\citenamefont {De~Lorenzo}\ and\ \citenamefont
  {Perez}(2016)}]{DeLorenzo2016}%
  \BibitemOpen
  \bibfield  {author} {\bibinfo {author} {\bibfnamefont {T.}~\bibnamefont
  {De~Lorenzo}}\ and\ \bibinfo {author} {\bibfnamefont {A.}~\bibnamefont
  {Perez}},\ }\bibfield  {title} {\bibinfo {title} {Improved black hole
  fireworks: Asymmetric black-hole-to-white-hole tunneling scenario},\
  }\bibfield  {journal} {\bibinfo  {journal} {Physical Review D}\ }\textbf
  {\bibinfo {volume} {93}},\ \href {https://doi.org/10.1103/physrevd.93.124018}
  {10.1103/physrevd.93.124018} (\bibinfo {year} {2016})\BibitemShut {NoStop}%
\bibitem [{\citenamefont {Husain}\ \emph
  {et~al.}(2022{\natexlab{a}})\citenamefont {Husain}, \citenamefont {Kelly},
  \citenamefont {Santacruz},\ and\ \citenamefont {Wilson-Ewing}}]{Husain2022}%
  \BibitemOpen
  \bibfield  {author} {\bibinfo {author} {\bibfnamefont {V.}~\bibnamefont
  {Husain}}, \bibinfo {author} {\bibfnamefont {J.~G.}\ \bibnamefont {Kelly}},
  \bibinfo {author} {\bibfnamefont {R.}~\bibnamefont {Santacruz}},\ and\
  \bibinfo {author} {\bibfnamefont {E.}~\bibnamefont {Wilson-Ewing}},\
  }\bibfield  {title} {\bibinfo {title} {Quantum gravity of dust collapse:
  Shock waves from black holes},\ }\bibfield  {journal} {\bibinfo  {journal}
  {Physical Review Letters}\ }\textbf {\bibinfo {volume} {128}},\ \href
  {https://doi.org/10.1103/physrevlett.128.121301}
  {10.1103/physrevlett.128.121301} (\bibinfo {year}
  {2022}{\natexlab{a}})\BibitemShut {NoStop}%
\bibitem [{\citenamefont {Achour}\ \emph
  {et~al.}(2020{\natexlab{b}})\citenamefont {Achour}, \citenamefont {Brahma},\
  and\ \citenamefont {Uzan}}]{Achour2020v1}%
  \BibitemOpen
  \bibfield  {author} {\bibinfo {author} {\bibfnamefont {J.~B.}\ \bibnamefont
  {Achour}}, \bibinfo {author} {\bibfnamefont {S.}~\bibnamefont {Brahma}},\
  and\ \bibinfo {author} {\bibfnamefont {J.-P.}\ \bibnamefont {Uzan}},\
  }\bibfield  {title} {\bibinfo {title} {Bouncing compact objects. part i.
  quantum extension of the oppenheimer-snyder collapse},\ }\href
  {https://doi.org/10.1088/1475-7516/2020/03/041} {\bibfield  {journal}
  {\bibinfo  {journal} {Journal of Cosmology and Astroparticle Physics}\
  }\textbf {\bibinfo {volume} {2020}}\bibinfo  {number} { (03)}}\BibitemShut
  {NoStop}%
\bibitem [{\citenamefont {Ben~Achour}\ and\ \citenamefont
  {Uzan}(2020)}]{BenAchour2020}%
  \BibitemOpen
\bibfield  {number} {  }\bibfield  {author} {\bibinfo {author} {\bibfnamefont
  {J.}~\bibnamefont {Ben~Achour}}\ and\ \bibinfo {author} {\bibfnamefont
  {J.-P.}\ \bibnamefont {Uzan}},\ }\bibfield  {title} {\bibinfo {title}
  {Bouncing compact objects. ii. effective theory of a pulsating planck star},\
  }\bibfield  {journal} {\bibinfo  {journal} {Physical Review D}\ }\textbf
  {\bibinfo {volume} {102}},\ \href
  {https://doi.org/10.1103/physrevd.102.124041} {10.1103/physrevd.102.124041}
  (\bibinfo {year} {2020})\BibitemShut {NoStop}%
\bibitem [{\citenamefont {Barca}\ and\ \citenamefont
  {Montani}(2024)}]{Barca2023}%
  \BibitemOpen
  \bibfield  {author} {\bibinfo {author} {\bibfnamefont {G.}~\bibnamefont
  {Barca}}\ and\ \bibinfo {author} {\bibfnamefont {G.}~\bibnamefont
  {Montani}},\ }\bibfield  {title} {\bibinfo {title} {{Non-singular
  gravitational collapse through modified Heisenberg algebra}},\ }\href
  {https://doi.org/10.1140/epjc/s10052-024-12564-5} {\bibfield  {journal}
  {\bibinfo  {journal} {Eur. Phys. J. C}\ }\textbf {\bibinfo {volume} {84}},\
  \bibinfo {pages} {261} (\bibinfo {year} {2024})}\BibitemShut {NoStop}%
\bibitem [{\citenamefont {Jusufi}(2023)}]{Jusufi2023}%
  \BibitemOpen
  \bibfield  {author} {\bibinfo {author} {\bibfnamefont {K.}~\bibnamefont
  {Jusufi}},\ }\bibfield  {title} {\bibinfo {title} {Avoidance of singularity
  during the gravitational collapse with string t-duality effects},\ }\bibfield
   {journal} {\bibinfo  {journal} {Universe}\ }\textbf {\bibinfo {volume}
  {9}},\ \href {https://doi.org/10.3390/universe9010041}
  {10.3390/universe9010041} (\bibinfo {year} {2023})\BibitemShut {NoStop}%
\bibitem [{\citenamefont {Fazzini}\ \emph {et~al.}(2023)\citenamefont
  {Fazzini}, \citenamefont {Rovelli},\ and\ \citenamefont
  {Soltani}}]{Fazzini2023}%
  \BibitemOpen
  \bibfield  {author} {\bibinfo {author} {\bibfnamefont {F.}~\bibnamefont
  {Fazzini}}, \bibinfo {author} {\bibfnamefont {C.}~\bibnamefont {Rovelli}},\
  and\ \bibinfo {author} {\bibfnamefont {F.}~\bibnamefont {Soltani}},\
  }\bibfield  {title} {\bibinfo {title} {Painlev\'e-gullstrand coordinates
  discontinuity in the quantum oppenheimer-snyder model},\ }\href
  {https://doi.org/10.1103/PhysRevD.108.044009} {\bibfield  {journal} {\bibinfo
   {journal} {Phys. Rev. D}\ }\textbf {\bibinfo {volume} {108}},\ \bibinfo
  {pages} {044009} (\bibinfo {year} {2023})}\BibitemShut {NoStop}%
\bibitem [{\citenamefont {Fazzini}\ \emph {et~al.}(2024)\citenamefont
  {Fazzini}, \citenamefont {Husain},\ and\ \citenamefont
  {Wilson-Ewing}}]{Fazzini2024}%
  \BibitemOpen
  \bibfield  {author} {\bibinfo {author} {\bibfnamefont {F.}~\bibnamefont
  {Fazzini}}, \bibinfo {author} {\bibfnamefont {V.}~\bibnamefont {Husain}},\
  and\ \bibinfo {author} {\bibfnamefont {E.}~\bibnamefont {Wilson-Ewing}},\
  }\bibfield  {title} {\bibinfo {title} {Shell-crossings and shock formation
  during gravitational collapse in effective loop quantum gravity},\ }\href
  {https://doi.org/10.1103/PhysRevD.109.084052} {\bibfield  {journal} {\bibinfo
   {journal} {Phys. Rev. D}\ }\textbf {\bibinfo {volume} {109}},\ \bibinfo
  {pages} {084052} (\bibinfo {year} {2024})}\BibitemShut {NoStop}%
\bibitem [{\citenamefont {Husain}\ \emph
  {et~al.}(2022{\natexlab{b}})\citenamefont {Husain}, \citenamefont {Kelly},
  \citenamefont {Santacruz},\ and\ \citenamefont
  {Wilson-Ewing}}]{HusainRev2022}%
  \BibitemOpen
  \bibfield  {author} {\bibinfo {author} {\bibfnamefont {V.}~\bibnamefont
  {Husain}}, \bibinfo {author} {\bibfnamefont {J.~G.}\ \bibnamefont {Kelly}},
  \bibinfo {author} {\bibfnamefont {R.}~\bibnamefont {Santacruz}},\ and\
  \bibinfo {author} {\bibfnamefont {E.}~\bibnamefont {Wilson-Ewing}},\
  }\bibfield  {title} {\bibinfo {title} {Fate of quantum black holes},\ }\href
  {https://doi.org/10.1103/PhysRevD.106.024014} {\bibfield  {journal} {\bibinfo
   {journal} {Phys. Rev. D}\ }\textbf {\bibinfo {volume} {106}},\ \bibinfo
  {pages} {024014} (\bibinfo {year} {2022}{\natexlab{b}})}\BibitemShut
  {NoStop}%
\bibitem [{\citenamefont {Bambi}\ \emph {et~al.}(2013)\citenamefont {Bambi},
  \citenamefont {Malafarina},\ and\ \citenamefont {Modesto}}]{Bambi2013}%
  \BibitemOpen
  \bibfield  {author} {\bibinfo {author} {\bibfnamefont {C.}~\bibnamefont
  {Bambi}}, \bibinfo {author} {\bibfnamefont {D.}~\bibnamefont {Malafarina}},\
  and\ \bibinfo {author} {\bibfnamefont {L.}~\bibnamefont {Modesto}},\
  }\bibfield  {title} {\bibinfo {title} {{Non-singular quantum-inspired
  gravitational collapse}},\ }\href
  {https://doi.org/10.1103/PhysRevD.88.044009} {\bibfield  {journal} {\bibinfo
  {journal} {Phys. Rev. D}\ }\textbf {\bibinfo {volume} {88}},\ \bibinfo
  {pages} {044009} (\bibinfo {year} {2013})}\BibitemShut {NoStop}%
\bibitem [{\citenamefont {Liu}\ \emph {et~al.}(2014)\citenamefont {Liu},
  \citenamefont {Malafarina}, \citenamefont {Modesto},\ and\ \citenamefont
  {Bambi}}]{Liu2014}%
  \BibitemOpen
  \bibfield  {author} {\bibinfo {author} {\bibfnamefont {Y.}~\bibnamefont
  {Liu}}, \bibinfo {author} {\bibfnamefont {D.}~\bibnamefont {Malafarina}},
  \bibinfo {author} {\bibfnamefont {L.}~\bibnamefont {Modesto}},\ and\ \bibinfo
  {author} {\bibfnamefont {C.}~\bibnamefont {Bambi}},\ }\bibfield  {title}
  {\bibinfo {title} {{Singularity avoidance in quantum-inspired inhomogeneous
  dust collapse}},\ }\href {https://doi.org/10.1103/PhysRevD.90.044040}
  {\bibfield  {journal} {\bibinfo  {journal} {Phys. Rev. D}\ }\textbf {\bibinfo
  {volume} {90}},\ \bibinfo {pages} {044040} (\bibinfo {year}
  {2014})}\BibitemShut {NoStop}%
\bibitem [{\citenamefont {Penrose}(1963)}]{Penrose1963}%
  \BibitemOpen
  \bibfield  {author} {\bibinfo {author} {\bibfnamefont {R.}~\bibnamefont
  {Penrose}},\ }\bibfield  {title} {\bibinfo {title} {Asymptotic properties of
  fields and space-times},\ }\bibfield  {journal} {\bibinfo  {journal}
  {Physical Review Letters}\ }\textbf {\bibinfo {volume} {10}},\ \href
  {https://doi.org/10.1103/physrevlett.10.66} {10.1103/physrevlett.10.66}
  (\bibinfo {year} {1963})\BibitemShut {NoStop}%
\bibitem [{\citenamefont {{Planck Collaboration}}\ \emph
  {et~al.}(2020)\citenamefont {{Planck Collaboration}}, \citenamefont
  {{Aghanim, N.}}, \citenamefont {{Akrami, Y.}}, \citenamefont {{Ashdown, M.}},
  \citenamefont {{Aumont, J.}}, \citenamefont {{Baccigalupi, C.}},
  \citenamefont {{Ballardini, M.}}, \citenamefont {{Banday, A. J.}},
  \citenamefont {{Barreiro, R. B.}}, \citenamefont {{Bartolo, N.}},
  \citenamefont {{Basak, S.}}, \citenamefont {{Battye, R.}}, \citenamefont
  {{Benabed, K.}}, \citenamefont {{Bernard, J.-P.}}, \citenamefont
  {{Bersanelli, M.}}, \citenamefont {{Bielewicz, P.}}, \citenamefont {{Bock, J.
  J.}}, \citenamefont {{Bond, J. R.}}, \citenamefont {{Borrill, J.}},
  \citenamefont {{Bouchet, F. R.}}, \citenamefont {{Boulanger, F.}},
  \citenamefont {{Bucher, M.}}, \citenamefont {{Burigana, C.}}, \citenamefont
  {{Butler, R. C.}}, \citenamefont {{Calabrese, E.}}, \citenamefont {{Cardoso,
  J.-F.}}, \citenamefont {{Carron, J.}}, \citenamefont {{Challinor, A.}},
  \citenamefont {{Chiang, H. C.}}, \citenamefont {{Chluba, J.}}, \citenamefont
  {{Colombo, L. P. L.}}, \citenamefont {{Combet, C.}}, \citenamefont
  {{Contreras, D.}}, \citenamefont {{Crill, B. P.}}, \citenamefont {{Cuttaia,
  F.}}, \citenamefont {{de Bernardis, P.}}, \citenamefont {{de Zotti, G.}},
  \citenamefont {{Delabrouille, J.}}, \citenamefont {{Delouis, J.-M.}},
  \citenamefont {{Di Valentino, E.}}, \citenamefont {{Diego, J. M.}},
  \citenamefont {{Doré, O.}}, \citenamefont {{Douspis, M.}}, \citenamefont
  {{Ducout, A.}}, \citenamefont {{Dupac, X.}}, \citenamefont {{Dusini, S.}},
  \citenamefont {{Efstathiou, G.}}, \citenamefont {{Elsner, F.}}, \citenamefont
  {{Enßlin, T. A.}}, \citenamefont {{Eriksen, H. K.}}, \citenamefont
  {{Fantaye, Y.}}, \citenamefont {{Farhang, M.}}, \citenamefont {{Fergusson,
  J.}}, \citenamefont {{Fernandez-Cobos, R.}}, \citenamefont {{Finelli, F.}},
  \citenamefont {{Forastieri, F.}}, \citenamefont {{Frailis, M.}},
  \citenamefont {{Fraisse, A. A.}}, \citenamefont {{Franceschi, E.}},
  \citenamefont {{Frolov, A.}}, \citenamefont {{Galeotta, S.}}, \citenamefont
  {{Galli, S.}}, \citenamefont {{Ganga, K.}}, \citenamefont {{Génova-Santos,
  R. T.}}, \citenamefont {{Gerbino, M.}}, \citenamefont {{Ghosh, T.}},
  \citenamefont {{González-Nuevo, J.}}, \citenamefont {{Górski, K. M.}},
  \citenamefont {{Gratton, S.}}, \citenamefont {{Gruppuso, A.}}, \citenamefont
  {{Gudmundsson, J. E.}}, \citenamefont {{Hamann, J.}}, \citenamefont
  {{Handley, W.}}, \citenamefont {{Hansen, F. K.}}, \citenamefont {{Herranz,
  D.}}, \citenamefont {{Hildebrandt, S. R.}}, \citenamefont {{Hivon, E.}},
  \citenamefont {{Huang, Z.}}, \citenamefont {{Jaffe, A. H.}}, \citenamefont
  {{Jones, W. C.}}, \citenamefont {{Karakci, A.}}, \citenamefont {{Keihänen,
  E.}}, \citenamefont {{Keskitalo, R.}}, \citenamefont {{Kiiveri, K.}},
  \citenamefont {{Kim, J.}}, \citenamefont {{Kisner, T. S.}}, \citenamefont
  {{Knox, L.}}, \citenamefont {{Krachmalnicoff, N.}}, \citenamefont {{Kunz,
  M.}}, \citenamefont {{Kurki-Suonio, H.}}, \citenamefont {{Lagache, G.}},
  \citenamefont {{Lamarre, J.-M.}}, \citenamefont {{Lasenby, A.}},
  \citenamefont {{Lattanzi, M.}}, \citenamefont {{Lawrence, C. R.}},
  \citenamefont {{Le Jeune, M.}}, \citenamefont {{Lemos, P.}}, \citenamefont
  {{Lesgourgues, J.}}, \citenamefont {{Levrier, F.}}, \citenamefont {{Lewis,
  A.}}, \citenamefont {{Liguori, M.}}, \citenamefont {{Lilje, P. B.}},
  \citenamefont {{Lilley, M.}}, \citenamefont {{Lindholm, V.}}, \citenamefont
  {{López-Caniego, M.}}, \citenamefont {{Lubin, P. M.}}, \citenamefont {{Ma,
  Y.-Z.}}, \citenamefont {{Macías-Pérez, J. F.}}, \citenamefont {{Maggio,
  G.}}, \citenamefont {{Maino, D.}}, \citenamefont {{Mandolesi, N.}},
  \citenamefont {{Mangilli, A.}}, \citenamefont {{Marcos-Caballero, A.}},
  \citenamefont {{Maris, M.}}, \citenamefont {{Martin, P. G.}}, \citenamefont
  {{Martinelli, M.}}, \citenamefont {{Martínez-González, E.}}, \citenamefont
  {{Matarrese, S.}}, \citenamefont {{Mauri, N.}}, \citenamefont {{McEwen, J.
  D.}}, \citenamefont {{Meinhold, P. R.}}, \citenamefont {{Melchiorri, A.}},
  \citenamefont {{Mennella, A.}}, \citenamefont {{Migliaccio, M.}},
  \citenamefont {{Millea, M.}}, \citenamefont {{Mitra, S.}}, \citenamefont
  {{Miville-Deschênes, M.-A.}}, \citenamefont {{Molinari, D.}}, \citenamefont
  {{Montier, L.}}, \citenamefont {{Morgante, G.}}, \citenamefont {{Moss, A.}},
  \citenamefont {{Natoli, P.}}, \citenamefont {{Nørgaard-Nielsen, H. U.}},
  \citenamefont {{Pagano, L.}}, \citenamefont {{Paoletti, D.}}, \citenamefont
  {{Partridge, B.}}, \citenamefont {{Patanchon, G.}}, \citenamefont {{Peiris,
  H. V.}}, \citenamefont {{Perrotta, F.}}, \citenamefont {{Pettorino, V.}},
  \citenamefont {{Piacentini, F.}}, \citenamefont {{Polastri, L.}},
  \citenamefont {{Polenta, G.}}, \citenamefont {{Puget, J.-L.}}, \citenamefont
  {{Rachen, J. P.}}, \citenamefont {{Reinecke, M.}}, \citenamefont
  {{Remazeilles, M.}}, \citenamefont {{Renzi, A.}}, \citenamefont {{Rocha,
  G.}}, \citenamefont {{Rosset, C.}}, \citenamefont {{Roudier, G.}},
  \citenamefont {{Rubiño-Martín, J. A.}}, \citenamefont {{Ruiz-Granados,
  B.}}, \citenamefont {{Salvati, L.}}, \citenamefont {{Sandri, M.}},
  \citenamefont {{Savelainen, M.}}, \citenamefont {{Scott, D.}}, \citenamefont
  {{Shellard, E. P. S.}}, \citenamefont {{Sirignano, C.}}, \citenamefont
  {{Sirri, G.}}, \citenamefont {{Spencer, L. D.}}, \citenamefont {{Sunyaev,
  R.}}, \citenamefont {{Suur-Uski, A.-S.}}, \citenamefont {{Tauber, J. A.}},
  \citenamefont {{Tavagnacco, D.}}, \citenamefont {{Tenti, M.}}, \citenamefont
  {{Toffolatti, L.}}, \citenamefont {{Tomasi, M.}}, \citenamefont {{Trombetti,
  T.}}, \citenamefont {{Valenziano, L.}}, \citenamefont {{Valiviita, J.}},
  \citenamefont {{Van Tent, B.}}, \citenamefont {{Vibert, L.}}, \citenamefont
  {{Vielva, P.}}, \citenamefont {{Villa, F.}}, \citenamefont {{Vittorio, N.}},
  \citenamefont {{Wandelt, B. D.}}, \citenamefont {{Wehus, I. K.}},
  \citenamefont {{White, M.}}, \citenamefont {{White, S. D. M.}}, \citenamefont
  {{Zacchei, A.}},\ and\ \citenamefont {{Zonca, A.}}}]{Planck2018}%
  \BibitemOpen
  \bibfield  {author} {\bibinfo {author} {\bibnamefont {{Planck
  Collaboration}}}, \bibinfo {author} {\bibnamefont {{Aghanim, N.}}}, \bibinfo
  {author} {\bibnamefont {{Akrami, Y.}}}, \bibinfo {author} {\bibnamefont
  {{Ashdown, M.}}}, \bibinfo {author} {\bibnamefont {{Aumont, J.}}}, \bibinfo
  {author} {\bibnamefont {{Baccigalupi, C.}}}, \bibinfo {author} {\bibnamefont
  {{Ballardini, M.}}}, \bibinfo {author} {\bibnamefont {{Banday, A. J.}}},
  \bibinfo {author} {\bibnamefont {{Barreiro, R. B.}}}, \bibinfo {author}
  {\bibnamefont {{Bartolo, N.}}}, \bibinfo {author} {\bibnamefont {{Basak,
  S.}}}, \bibinfo {author} {\bibnamefont {{Battye, R.}}}, \bibinfo {author}
  {\bibnamefont {{Benabed, K.}}}, \bibinfo {author} {\bibnamefont {{Bernard,
  J.-P.}}}, \bibinfo {author} {\bibnamefont {{Bersanelli, M.}}}, \bibinfo
  {author} {\bibnamefont {{Bielewicz, P.}}}, \bibinfo {author} {\bibnamefont
  {{Bock, J. J.}}}, \bibinfo {author} {\bibnamefont {{Bond, J. R.}}}, \bibinfo
  {author} {\bibnamefont {{Borrill, J.}}}, \bibinfo {author} {\bibnamefont
  {{Bouchet, F. R.}}}, \bibinfo {author} {\bibnamefont {{Boulanger, F.}}},
  \bibinfo {author} {\bibnamefont {{Bucher, M.}}}, \bibinfo {author}
  {\bibnamefont {{Burigana, C.}}}, \bibinfo {author} {\bibnamefont {{Butler, R.
  C.}}}, \bibinfo {author} {\bibnamefont {{Calabrese, E.}}}, \bibinfo {author}
  {\bibnamefont {{Cardoso, J.-F.}}}, \bibinfo {author} {\bibnamefont {{Carron,
  J.}}}, \bibinfo {author} {\bibnamefont {{Challinor, A.}}}, \bibinfo {author}
  {\bibnamefont {{Chiang, H. C.}}}, \bibinfo {author} {\bibnamefont {{Chluba,
  J.}}}, \bibinfo {author} {\bibnamefont {{Colombo, L. P. L.}}}, \bibinfo
  {author} {\bibnamefont {{Combet, C.}}}, \bibinfo {author} {\bibnamefont
  {{Contreras, D.}}}, \bibinfo {author} {\bibnamefont {{Crill, B. P.}}},
  \bibinfo {author} {\bibnamefont {{Cuttaia, F.}}}, \bibinfo {author}
  {\bibnamefont {{de Bernardis, P.}}}, \bibinfo {author} {\bibnamefont {{de
  Zotti, G.}}}, \bibinfo {author} {\bibnamefont {{Delabrouille, J.}}}, \bibinfo
  {author} {\bibnamefont {{Delouis, J.-M.}}}, \bibinfo {author} {\bibnamefont
  {{Di Valentino, E.}}}, \bibinfo {author} {\bibnamefont {{Diego, J. M.}}},
  \bibinfo {author} {\bibnamefont {{Doré, O.}}}, \bibinfo {author}
  {\bibnamefont {{Douspis, M.}}}, \bibinfo {author} {\bibnamefont {{Ducout,
  A.}}}, \bibinfo {author} {\bibnamefont {{Dupac, X.}}}, \bibinfo {author}
  {\bibnamefont {{Dusini, S.}}}, \bibinfo {author} {\bibnamefont {{Efstathiou,
  G.}}}, \bibinfo {author} {\bibnamefont {{Elsner, F.}}}, \bibinfo {author}
  {\bibnamefont {{Enßlin, T. A.}}}, \bibinfo {author} {\bibnamefont {{Eriksen,
  H. K.}}}, \bibinfo {author} {\bibnamefont {{Fantaye, Y.}}}, \bibinfo {author}
  {\bibnamefont {{Farhang, M.}}}, \bibinfo {author} {\bibnamefont {{Fergusson,
  J.}}}, \bibinfo {author} {\bibnamefont {{Fernandez-Cobos, R.}}}, \bibinfo
  {author} {\bibnamefont {{Finelli, F.}}}, \bibinfo {author} {\bibnamefont
  {{Forastieri, F.}}}, \bibinfo {author} {\bibnamefont {{Frailis, M.}}},
  \bibinfo {author} {\bibnamefont {{Fraisse, A. A.}}}, \bibinfo {author}
  {\bibnamefont {{Franceschi, E.}}}, \bibinfo {author} {\bibnamefont {{Frolov,
  A.}}}, \bibinfo {author} {\bibnamefont {{Galeotta, S.}}}, \bibinfo {author}
  {\bibnamefont {{Galli, S.}}}, \bibinfo {author} {\bibnamefont {{Ganga, K.}}},
  \bibinfo {author} {\bibnamefont {{Génova-Santos, R. T.}}}, \bibinfo {author}
  {\bibnamefont {{Gerbino, M.}}}, \bibinfo {author} {\bibnamefont {{Ghosh,
  T.}}}, \bibinfo {author} {\bibnamefont {{González-Nuevo, J.}}}, \bibinfo
  {author} {\bibnamefont {{Górski, K. M.}}}, \bibinfo {author} {\bibnamefont
  {{Gratton, S.}}}, \bibinfo {author} {\bibnamefont {{Gruppuso, A.}}}, \bibinfo
  {author} {\bibnamefont {{Gudmundsson, J. E.}}}, \bibinfo {author}
  {\bibnamefont {{Hamann, J.}}}, \bibinfo {author} {\bibnamefont {{Handley,
  W.}}}, \bibinfo {author} {\bibnamefont {{Hansen, F. K.}}}, \bibinfo {author}
  {\bibnamefont {{Herranz, D.}}}, \bibinfo {author} {\bibnamefont
  {{Hildebrandt, S. R.}}}, \bibinfo {author} {\bibnamefont {{Hivon, E.}}},
  \bibinfo {author} {\bibnamefont {{Huang, Z.}}}, \bibinfo {author}
  {\bibnamefont {{Jaffe, A. H.}}}, \bibinfo {author} {\bibnamefont {{Jones, W.
  C.}}}, \bibinfo {author} {\bibnamefont {{Karakci, A.}}}, \bibinfo {author}
  {\bibnamefont {{Keihänen, E.}}}, \bibinfo {author} {\bibnamefont
  {{Keskitalo, R.}}}, \bibinfo {author} {\bibnamefont {{Kiiveri, K.}}},
  \bibinfo {author} {\bibnamefont {{Kim, J.}}}, \bibinfo {author} {\bibnamefont
  {{Kisner, T. S.}}}, \bibinfo {author} {\bibnamefont {{Knox, L.}}}, \bibinfo
  {author} {\bibnamefont {{Krachmalnicoff, N.}}}, \bibinfo {author}
  {\bibnamefont {{Kunz, M.}}}, \bibinfo {author} {\bibnamefont {{Kurki-Suonio,
  H.}}}, \bibinfo {author} {\bibnamefont {{Lagache, G.}}}, \bibinfo {author}
  {\bibnamefont {{Lamarre, J.-M.}}}, \bibinfo {author} {\bibnamefont {{Lasenby,
  A.}}}, \bibinfo {author} {\bibnamefont {{Lattanzi, M.}}}, \bibinfo {author}
  {\bibnamefont {{Lawrence, C. R.}}}, \bibinfo {author} {\bibnamefont {{Le
  Jeune, M.}}}, \bibinfo {author} {\bibnamefont {{Lemos, P.}}}, \bibinfo
  {author} {\bibnamefont {{Lesgourgues, J.}}}, \bibinfo {author} {\bibnamefont
  {{Levrier, F.}}}, \bibinfo {author} {\bibnamefont {{Lewis, A.}}}, \bibinfo
  {author} {\bibnamefont {{Liguori, M.}}}, \bibinfo {author} {\bibnamefont
  {{Lilje, P. B.}}}, \bibinfo {author} {\bibnamefont {{Lilley, M.}}}, \bibinfo
  {author} {\bibnamefont {{Lindholm, V.}}}, \bibinfo {author} {\bibnamefont
  {{López-Caniego, M.}}}, \bibinfo {author} {\bibnamefont {{Lubin, P. M.}}},
  \bibinfo {author} {\bibnamefont {{Ma, Y.-Z.}}}, \bibinfo {author}
  {\bibnamefont {{Macías-Pérez, J. F.}}}, \bibinfo {author} {\bibnamefont
  {{Maggio, G.}}}, \bibinfo {author} {\bibnamefont {{Maino, D.}}}, \bibinfo
  {author} {\bibnamefont {{Mandolesi, N.}}}, \bibinfo {author} {\bibnamefont
  {{Mangilli, A.}}}, \bibinfo {author} {\bibnamefont {{Marcos-Caballero, A.}}},
  \bibinfo {author} {\bibnamefont {{Maris, M.}}}, \bibinfo {author}
  {\bibnamefont {{Martin, P. G.}}}, \bibinfo {author} {\bibnamefont
  {{Martinelli, M.}}}, \bibinfo {author} {\bibnamefont {{Martínez-González,
  E.}}}, \bibinfo {author} {\bibnamefont {{Matarrese, S.}}}, \bibinfo {author}
  {\bibnamefont {{Mauri, N.}}}, \bibinfo {author} {\bibnamefont {{McEwen, J.
  D.}}}, \bibinfo {author} {\bibnamefont {{Meinhold, P. R.}}}, \bibinfo
  {author} {\bibnamefont {{Melchiorri, A.}}}, \bibinfo {author} {\bibnamefont
  {{Mennella, A.}}}, \bibinfo {author} {\bibnamefont {{Migliaccio, M.}}},
  \bibinfo {author} {\bibnamefont {{Millea, M.}}}, \bibinfo {author}
  {\bibnamefont {{Mitra, S.}}}, \bibinfo {author} {\bibnamefont
  {{Miville-Deschênes, M.-A.}}}, \bibinfo {author} {\bibnamefont {{Molinari,
  D.}}}, \bibinfo {author} {\bibnamefont {{Montier, L.}}}, \bibinfo {author}
  {\bibnamefont {{Morgante, G.}}}, \bibinfo {author} {\bibnamefont {{Moss,
  A.}}}, \bibinfo {author} {\bibnamefont {{Natoli, P.}}}, \bibinfo {author}
  {\bibnamefont {{Nørgaard-Nielsen, H. U.}}}, \bibinfo {author} {\bibnamefont
  {{Pagano, L.}}}, \bibinfo {author} {\bibnamefont {{Paoletti, D.}}}, \bibinfo
  {author} {\bibnamefont {{Partridge, B.}}}, \bibinfo {author} {\bibnamefont
  {{Patanchon, G.}}}, \bibinfo {author} {\bibnamefont {{Peiris, H. V.}}},
  \bibinfo {author} {\bibnamefont {{Perrotta, F.}}}, \bibinfo {author}
  {\bibnamefont {{Pettorino, V.}}}, \bibinfo {author} {\bibnamefont
  {{Piacentini, F.}}}, \bibinfo {author} {\bibnamefont {{Polastri, L.}}},
  \bibinfo {author} {\bibnamefont {{Polenta, G.}}}, \bibinfo {author}
  {\bibnamefont {{Puget, J.-L.}}}, \bibinfo {author} {\bibnamefont {{Rachen, J.
  P.}}}, \bibinfo {author} {\bibnamefont {{Reinecke, M.}}}, \bibinfo {author}
  {\bibnamefont {{Remazeilles, M.}}}, \bibinfo {author} {\bibnamefont {{Renzi,
  A.}}}, \bibinfo {author} {\bibnamefont {{Rocha, G.}}}, \bibinfo {author}
  {\bibnamefont {{Rosset, C.}}}, \bibinfo {author} {\bibnamefont {{Roudier,
  G.}}}, \bibinfo {author} {\bibnamefont {{Rubiño-Martín, J. A.}}}, \bibinfo
  {author} {\bibnamefont {{Ruiz-Granados, B.}}}, \bibinfo {author}
  {\bibnamefont {{Salvati, L.}}}, \bibinfo {author} {\bibnamefont {{Sandri,
  M.}}}, \bibinfo {author} {\bibnamefont {{Savelainen, M.}}}, \bibinfo {author}
  {\bibnamefont {{Scott, D.}}}, \bibinfo {author} {\bibnamefont {{Shellard, E.
  P. S.}}}, \bibinfo {author} {\bibnamefont {{Sirignano, C.}}}, \bibinfo
  {author} {\bibnamefont {{Sirri, G.}}}, \bibinfo {author} {\bibnamefont
  {{Spencer, L. D.}}}, \bibinfo {author} {\bibnamefont {{Sunyaev, R.}}},
  \bibinfo {author} {\bibnamefont {{Suur-Uski, A.-S.}}}, \bibinfo {author}
  {\bibnamefont {{Tauber, J. A.}}}, \bibinfo {author} {\bibnamefont
  {{Tavagnacco, D.}}}, \bibinfo {author} {\bibnamefont {{Tenti, M.}}}, \bibinfo
  {author} {\bibnamefont {{Toffolatti, L.}}}, \bibinfo {author} {\bibnamefont
  {{Tomasi, M.}}}, \bibinfo {author} {\bibnamefont {{Trombetti, T.}}}, \bibinfo
  {author} {\bibnamefont {{Valenziano, L.}}}, \bibinfo {author} {\bibnamefont
  {{Valiviita, J.}}}, \bibinfo {author} {\bibnamefont {{Van Tent, B.}}},
  \bibinfo {author} {\bibnamefont {{Vibert, L.}}}, \bibinfo {author}
  {\bibnamefont {{Vielva, P.}}}, \bibinfo {author} {\bibnamefont {{Villa,
  F.}}}, \bibinfo {author} {\bibnamefont {{Vittorio, N.}}}, \bibinfo {author}
  {\bibnamefont {{Wandelt, B. D.}}}, \bibinfo {author} {\bibnamefont {{Wehus,
  I. K.}}}, \bibinfo {author} {\bibnamefont {{White, M.}}}, \bibinfo {author}
  {\bibnamefont {{White, S. D. M.}}}, \bibinfo {author} {\bibnamefont
  {{Zacchei, A.}}},\ and\ \bibinfo {author} {\bibnamefont {{Zonca, A.}}},\
  }\bibfield  {title} {\bibinfo {title} {Planck 2018 results - vi. cosmological
  parameters},\ }\href {https://doi.org/10.1051/0004-6361/201833910} {\bibfield
   {journal} {\bibinfo  {journal} {A\&A}\ }\textbf {\bibinfo {volume} {641}},\
  \bibinfo {pages} {A6} (\bibinfo {year} {2020})}\BibitemShut {NoStop}%
\end{thebibliography}%

\end{document}